\documentclass[preprint2]{aastex}

\slugcomment{Accepted for publication in the AJ; submitted 2021-08-17, accepted 2021-09-01.}

\shorttitle{Variability in PPNe: VIII. New SH Objects}

\shortauthors{Hrivnak et al.}

\begin{document}

\title{Variability in Proto-Planetary Nebulae: VIII. A New Sample of Southern Hemisphere Objects} 

\author{Bruce J. Hrivnak\altaffilmark{1,2}, Gary Henson\altaffilmark{2,3}, Todd C. Hillwig\altaffilmark{1,2}, Wenxian Lu\altaffilmark{1}, Matthew T. Bremer\altaffilmark{1}, David M. Vogl\altaffilmark{1}, Peyton J. Grimm\altaffilmark{1}, and Sean M. Egan\altaffilmark{1} }

\altaffiltext{1}{Department of Physics and Astronomy, Valparaiso University, 
Valparaiso, IN 46383, USA; bruce.hrivnak@valpo.edu, todd.hillwig@valpo.edu, wen.lu@valpo.edu (retired)}
\altaffiltext{2}{Southeastern Association for Research in Astronomy (SARA), USA}
\altaffiltext{3}{Department of Physics and Astronomy, East Tennessee State University, 
Johnson City, TN 37614, USA; hensong@mail.etsu.edu}

\begin{abstract}
As part of our continuing study of light variability in proto-planetary nebulae (PPNe), we present the results from a long-term study of nine southern hemisphere objects.  
We have monitored their light variations over a nine-year interval from 2010-2018.  These were supplemented by data from the ASAS-SN and ASAS-3 surveys, leading to combined light curves from 2000 to 2020.  Pulsation periods were found in seven of the objects, although the three shortest must be regarded as tentative.
The periods range from 24 to 73 days.  When compared with the results of previous studies of the light variations in PPNe, we find that they show the same trends of shorter period and smaller light variations with higher temperatures.
Luminosities were calculated based on the spectral energy distributions, reddening, and Gaia distances, and these confirm the identification of all but one as post-AGB objects.  
Three of the stars possess long-period variations of 5 to 19 years.  These are most likely due to the periodic obscuration of the star by a disk, suggesting the presence of a binary companion and a circumbinary disk. \\
\end{abstract}

\section{Introduction}

Proto-planetary nebulae\footnote{Also referred to as preplanetary nebulae; cf. \citet{sah05}.} (PPNe) are objects in transition between asymptotic giant branch (AGB) stars and planetary nebulae (PNe).  This transition happens relatively quickly, on the order of a few thousand years, depending upon the mass of the star \citep{milb16}.  That short timescale and the obscuration by circumstellar dust in the ejected envelope made it difficult to identify objects in this evolutionary state.
However, following the successful mission of the InfraRed Astronomy Satellite (IRAS) in 1983, a number of candidates have been identified based on their large excesses of radiation in the mid-infrared, peaking particularly in the 25 $\mu$m bandpass \citep{partha86,veen89,hri89}.  The following decades have witnessed a rapid expansion in our knowledge of these objects based on studies of the chemistry of the central stars \citep{vanwin00,rao12,des16} and the chemistry \citep{mol02,hri09}, morphology \citep{ueta00,sah07,siod08}, and kinematics of the surrounding circumstellar envelope \citep{buj01}.  

Longer-term studies of the central stars of PPNe have also been carried out to search for variability due to binarity \citep{hri17} and pulsation \citep{ark10,ark11,hri10,hri15}.
From studies, mostly of northern hemisphere objects, pulsation periods ranging from 35 to 160 days have been found, primarily based on the analysis of stars of spectral types F$-$G.  Hotter stars appear to vary on much shorter timescales of days \citep[e.g.,][]{ark13}.  

In this paper, we report on the study of a new sample of southern hemisphere PPNe.  After introducing the sample, we describe the observations we made and the additional data sets we used, the variability in light curves, and the results of the period analyses. Finally, we discuss these results in the context of the bigger picture of the study of pulsation in PPNe.

\section{Program Objects}

The program objects are listed in Table~\ref{object_list}, where we have identified them based upon their IRAS and 2MASS catalog numbers and also their catalog numbers in the extensive observational study of post-AGB candidates by \citet{garlar97}.  Also listed are their locations in the sky (equatorial and galactic coordinates), and their apparent brightnesses, colors, and spectral types.  The average {\it V} brightness ranges from 11.8 to 13.6 mag and their spectral types range from A3 to G0, with a supergiant luminosity classification.  Their ({\it B$-$V}) colors range from 0.8 to 1.6 mags, much redder than expected for these spectral types but consistent with the presence of significant interstellar and circumstellar reddening.
We note that IRAS 15310$-$6149 has been classified as both a Herbig Ae/Be star \citep{vie03} and a post-AGB star \citep{szcz07}.  A subsequent study has suggested that it is most likely a post-AGB star \citep{vie11}. 

\placetable{object_list} 

Although they possess circumstellar envelopes, we know very little about their morphologies.  
Only IRAS 08143$-$4406 has been imaged with the Hubble Space Telescope.
It is elliptical with a size of 2.1$\arcsec$$\times$1.4$\arcsec$ with a bright central star \citep{siod08}.
IRAS 14429$-$4539 has been imaged in the mid-infrared but was unresolved at 0.3$\arcsec$ \citep{lag11}.

\section{Observations $\&$ Data Sets}

We have used three data sets in this study.  
The first of these is composed of our observations made at Cerro-Tololo Interamerican Observatory.  
The other two are from on-line databases.

\subsection{Our SARA Observations}

Our observations were carried out remotely from May 2010 to Apr 2018 using the 0.6 m Lowell telescope at Cerro-Tololo Interamerican Observatory, operated by the Southeast Association for Research in Astronomy (SARA).  During this time, three different CCD detectors were used for imaging, each equipped with standard Johnson {\it V} and Cousins {\it R}$_C$ filters \citep{keel17}.   
The cameras were used for one to three successive seasons, without overlap, but with a gap in observations from 2011 August to 2013 April.  For each object, typically 15$-$30 observations with each filter were made with each camera.
We carried out a program of differential photometry, to allow us to observe in slightly non-photometric conditions.  
The observations were reduced using IRAF\footnote{IRAF is distributed by the National Optical Astronomical Observatory, operated by the Association for Universities for Research in Astronomy, Inc., under contract with the National Science Foundation.},
and involved removing cosmic rays, subtracting the bias, and flat fielding the images. 
Brightness measurements were made using aperture photometry, with an aperture of $\sim$5.5$\arcsec$ radius.
Landolt standard stars \citep{land83,land92} were observed on several nights with each of the detector systems, and the data were reduced using a slightly larger aperture of 7$\arcsec$ radius, as used by Landolt.  From these, we determined standard linear color coefficients, which were used in the transformation from the instrumental to the standard photometric system.  
The standard magnitudes and colors of the program stars are listed in Table~\ref{std_ppn}, along with the dates of observation.
Three comparison stars were used in each field, and they are identified and their standard magnitudes and colors listed in Table~\ref{std_comp}.
These comparison stars were found to be constant at the level of $\pm$0.01 to $\pm$0.02 mag for each camera system.

\placetable{std_ppn}

\placetable{std_comp} 

\begin{figure}\figurenum{1}\epsscale{1.8} 
\plotone{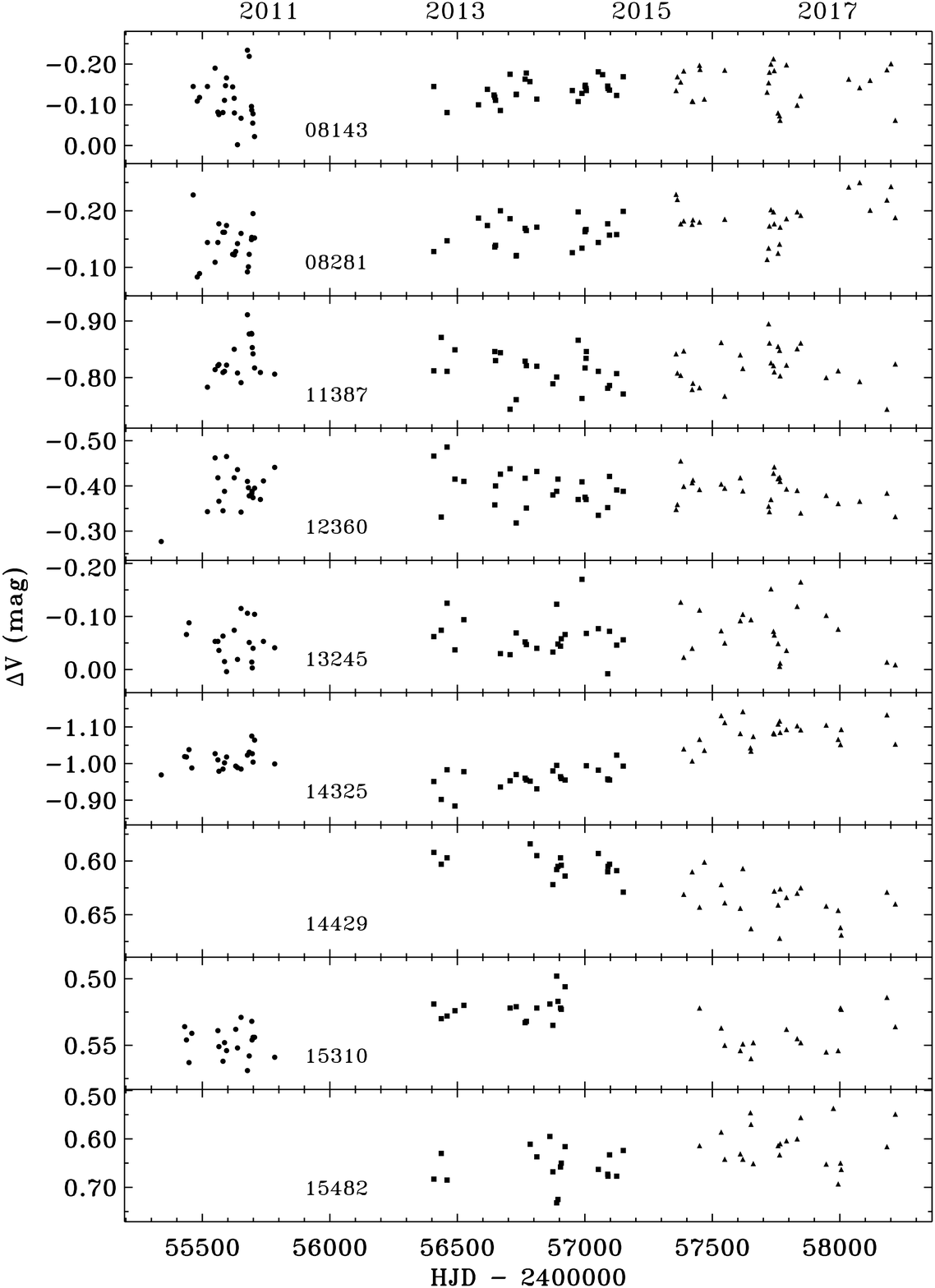}
\caption{The differential {\it V} SARA light curves from 2010$-$2018 for the nine program objects.  The observations with the three different cameras are shown with different symbols.
Typical error bars are $\pm$0.010 mag and are shown in Figure~\ref{LC_fits}. 
Note that the brightness ranges are smaller for IRAS 14429$-$4539 and 15310$-$6149.
\label{fig1}}
\epsscale{1.0}
\end{figure}

The use of  different detectors can result in some complications in the combined light curves for very reddened stars, as we found in our previous two studies in this series \citep{hri20a,hri20b}.  This was particularly the case for camera 1 of the present study.
However, this is much less problematic with the stars in the present study, which, though reddened, are not by nearly as much as the stars in the two previous studies.
For a few of the present stars, we comment upon possible small systematic offsets between the data from the different cameras, and in these cases we also analyzed 
the data with the light from the different cameras normalized to to the same level, and found similar results.
The combined SARA light curves for the {\it V} filter are shown in Figure~\ref{fig1}. 
The variations in light are not large, typically on the order of 0.1 mag in a season and range from about 0.1 to 0.25 mag over the entire eight year observing interval.
Two of the stars, IRAS 14429$-$4539 and 15482$-$5741, were added later to the program and observations of them began only in 2013 April and, consequently, only used cameras 2 and 3.
Uncertainties in the data are $\pm$0.010 mag.

\placefigure{fig1}

\begin{figure}\figurenum{2}\epsscale{1.0} 
\plotone{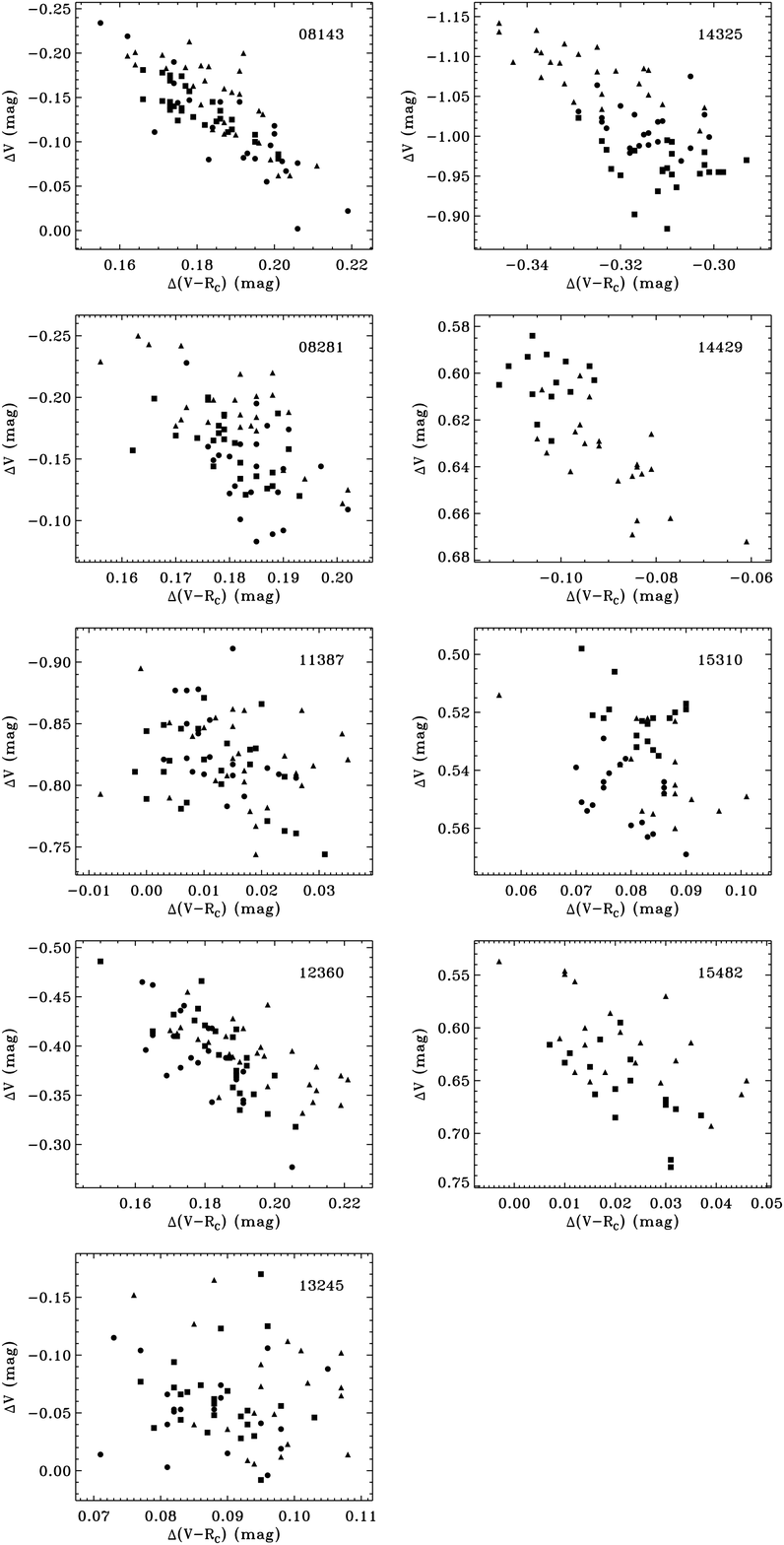}
\caption{The differential color curves for the target objects based on our new SARA data: $\Delta$({\it V$-$R$_C$}) versus $\Delta${\it V}.  There exists a general trend of the stars appearing redder when fainter, but it is not seen in all cases.  Symbols as in Figure~\ref{fig1}.
\label{color}}
\epsscale{1.0}
\end{figure}

The stars vary slightly in color, over a range in ({\it V$-$R}$_C$) of 0.03 to 0.06 mag.  
There exists a general tendency to be redder when fainter, but not in all cases.  This is shown in Figure~\ref{color}.
Any slight offset in magnitude or color between the data from the different cameras could show up here, but the general tendency to be redder when fainter appears to be solid.  
This trend has been seen in most other PPNe observed, although in previous studies which included PPNe of G spectral type, the color variations were larger. 
Thus the stars are cooler when fainter, with the exceptions of IRAS 13245$-$5036 and 15310$-$6149.
Three of the stars, IRAS 11387$-$6113, 13245$-$5036, and 14429$-$4539 have emission lines, particularly H$\alpha$, and that will affect the {\it R}$_C$ value and the {\it V$-$R}$_C$ color.

\placefigure{color}

\subsection{ASAS-SN Data}
\label{asas-sn-data}

Photometric measurements of each of these PPNe are also available from the All-Sky Automated Survey for Supernovae \citep[ASAS-SN;][]{koch17}.  
The survey uses several cameras on five different good astronomical sites in acquiring its data.  Observations have been made in two filters, {\it V} and {\it g} (the Sloan Digital Sky Survey {\it g} filter, $\lambda$$_{\rm eff}$=477 nm).  Exposure times are 90 sec, and typically the cameras make three successive dithered images.  
Stellar magnitudes were determined using aperture photometry with a radius of 16$\arcsec$, which is rather large.
We decided to use only the higher quality images, thus eliminating data with large full-width half maximum (FWHM) or large standard deviations.  Following the procedure of \citet{jay18}, we removed data with FWHM $>$ 2.0.  Then we eliminated data with large standard deviations and combined the successive observations into a single average nightly observation, keeping only data with average standard deviations of less than or equal to 0.015, 0.020, or 0.025 mag, depending upon the brightness of the object.  In most cases, especially for the {\it g} filter observations, objects were observed with several cameras.  In many cases, small calibration offsets exist between the observations with different cameras.
\citet{jay18} corrected for this by designating the camera with the largest number of observations as the primary, and then determined the needed additive offset to adjust the others to the primary by comparing the median values of the observations with each camera.
Since our objects are variable and different cameras are often used at different times of the observing interval, we adopted a different approach to determining the offsets.  We compared observations made on the same or adjacent nights by the primary and each of the secondary cameras, and from these determined the offsets.  In some cases, we used a boot-strapping method, in which we compared the offset between two secondary cameras to get the offset from the primary.  
The offsets ranged from $-$0.035 to +0.030, but in most cases were small, $-$0.005 to +0.005.  
Data from a particular camera were not used if they were only a small number of observations or if no good offset could be determined for that star.
The average uncertainties in the data range from $\pm$0.010$-$0.015 mag.

{\it V} observations for these stars are available from 2016 February through 2018 October, and thus cover about 2.5 seasons.  The {\it g} observations began in 2018 June and continue to the present time; we downloaded and used observations through 2020 June.  Thus the {\it g} data cover one season completely and parts of two others.  There is only a small overlap in which data are available contemporaneously in both filters.
In total, these observations cover approximately 4.5 observing seasons with high density coverage.
These ASAS-SN light curves are displayed in Figure~\ref{fig3}.  
To show both the {\it V} and {\it g} data easily on the same graph, we have added an empirically-determined offset to the {\it g} data to bring it to the brightness level of the {\it V} data, and then distinguished the two data sets by different colors and symbols.  This allows us to see the entire pattern of variability in a single figure.  Based on previous studies of PPNe light variability, we expect pulsation to be closely in phase in the {\it V} and {\it g} wavelength bands, but with the the pulsation amplitude larger in the shorter wavelength {\it g} band \citep{hri18}. 
About half of the stars show visible evidence of cyclical variability.

\placefigure{fig3}

\subsection{ASAS-3 Data}
\label{asas-data}

Photometric data for each of these objects were obtained in the All Sky Automated Survey, version 3 (ASAS-3)\footnote{http://www.astrouw.edu.pl/asas/} in the {\it V} filter from 2000$-$2009 (Nov 2000 to Nov 2009).  
Although the photometric uncertainties are rather large for these observations, typically $\pm$0.04$-$0.05 mag, they are useful for investigating trends in the data and perhaps even for determining or confirming periods in these objects.  To reduce contamination by nearby stars, we chose to use the data measured with an aperture of 15$\arcsec$ in radius.

We chose to use only the highest quality data (what are classified as Grade=A in the survey) and then further reduced the data to eliminate observations with larger uncertainties.  Through this procedure, we kept only data with uncertainties $\le$ 0.050 mag, except in the cases of IRAS 11387$-$6113 ($\le$ 0.065 mag) and 15482$-$5741 ($\le$ 0.055 mag).  A few of the objects happened to fall in regions of overlap between different telescope settings and thus had two or three large data sets, which we combined.  
We determined offsets, if they existed, by comparing observations made on the same or adjacent nights, at the different telescope settings.  
These empirically-determined offsets were generally small, but they were applied to bring all the data for a particular object to the level of the dataset with the largest number of observations.
A few very discrepant points were removed from the datasets.  
Average uncertainties in the data are $\pm$0.04$-$0.05 mag, except for IRAS 11387$-$6113, for which they are $\pm$0.06 mag.
We do not here display all of these light curves, but later we show three of them that are useful in establishing long-term, multi-year variability.

\section{Light Curves and Variability Study}

\subsection{Analysis Strategy}
\label{analysis}

We have several data sets at our disposal to examine and analyze for each object.  The following strategy was  adopted.
We began with our SARA {\it V} and {\it R}$_C$ light curves, which have the virtue of extending over a nine year interval with good precision.  
However, the observations are not numerous, 39$-$80 per filter for each object.
We began the analysis with the observed light curves and also analyzed them with the data from each camera normalized to the same level, to remove any small camera offsets.
The results were found to be similar between the observed and adjusted data sets.
Next the ASAS-SN {\it V} and {\it g} data were investigated.  They are of shorter intervals of time but with much higher density.  The {\it V} light curve covered most of three seasons, with $\sim$200 observations for each star.  The {\it g} data cover one full season and varying parts of two others, with some initial observations that overlap with the {\it V} light curve, and $\sim$300 observations for each star.  
(The one exception is IRAS 15310$-$6149, with 453 ({\it V}) and 591 ({\it g}) observations, as it was observed by several cameras.)
These ASAS-SN {\it V} and {\it g} data were also combined, with an empirically-determined offset based on the season of overlap (that brought the {\it g} data to the brightness level of the {\it V} data), and these combined light curves were analyzed.
While the amplitude of the light curve is expected to be larger in the {\it g} band, the data in the two filters are expected to be in phase \citep{hri18}, and thus an accurate period can be expected to be determined from these five years of high density, high precision data.
We then investigated a combined {\it V} light curve from the SARA and ASAS-SN observations, which was our preferred data set. 
The older ASAS-3 light curves were examined for long-term trends and periodicity.
In three cases 
where there was evidence of long-term variability, we also analyzed a {\it V} light curve composed of the combined ASAS-3, SARA, and ASAS-SN data (2000$-$2018).
For three of the stars, IRAS 14429$-$4539, 15310$-$6149, and 15482$-$5741, there was evidence of contamination of the ASAS-SN and ASAS-3 light curves by light from a relatively nearby bright star falling within the relatively large apertures used for those measurements.  This compromised the use of these data.
 Finally, for each star, we made an overall assessment of the evidence for periodic variability.  

The light curves are analyzed using PERIOD04 \citep{lenz05}.  This is a commonly used program for finding the periods in pulsating stars and has the advantage that it can fit multiple periods simultaneously.  It uses a Fourier analysis to find the most dominant frequency in the data and then fits the light curve with a sine curve, which works well in these low-amplitude pulsators.  Uncertainties in the data were determined by a least-squares analysis and also by a Monte Carlo analysis, and they gave similar results.  The results quoted are from the use of this program.  A period is judged to be significant if it meets the commonly used criterion of a signal-to-noise ratio (S/N) $\ge$ 4.0 \citep{bre93}. 
We also analyzed the light curves with several other commonly used period-search programs \citep[using Peranso;][]{paun16} to determine the dominant period and obtained similar results.

\begin{figure}\figurenum{4}\epsscale{1.0} 
\rotate
\plotone{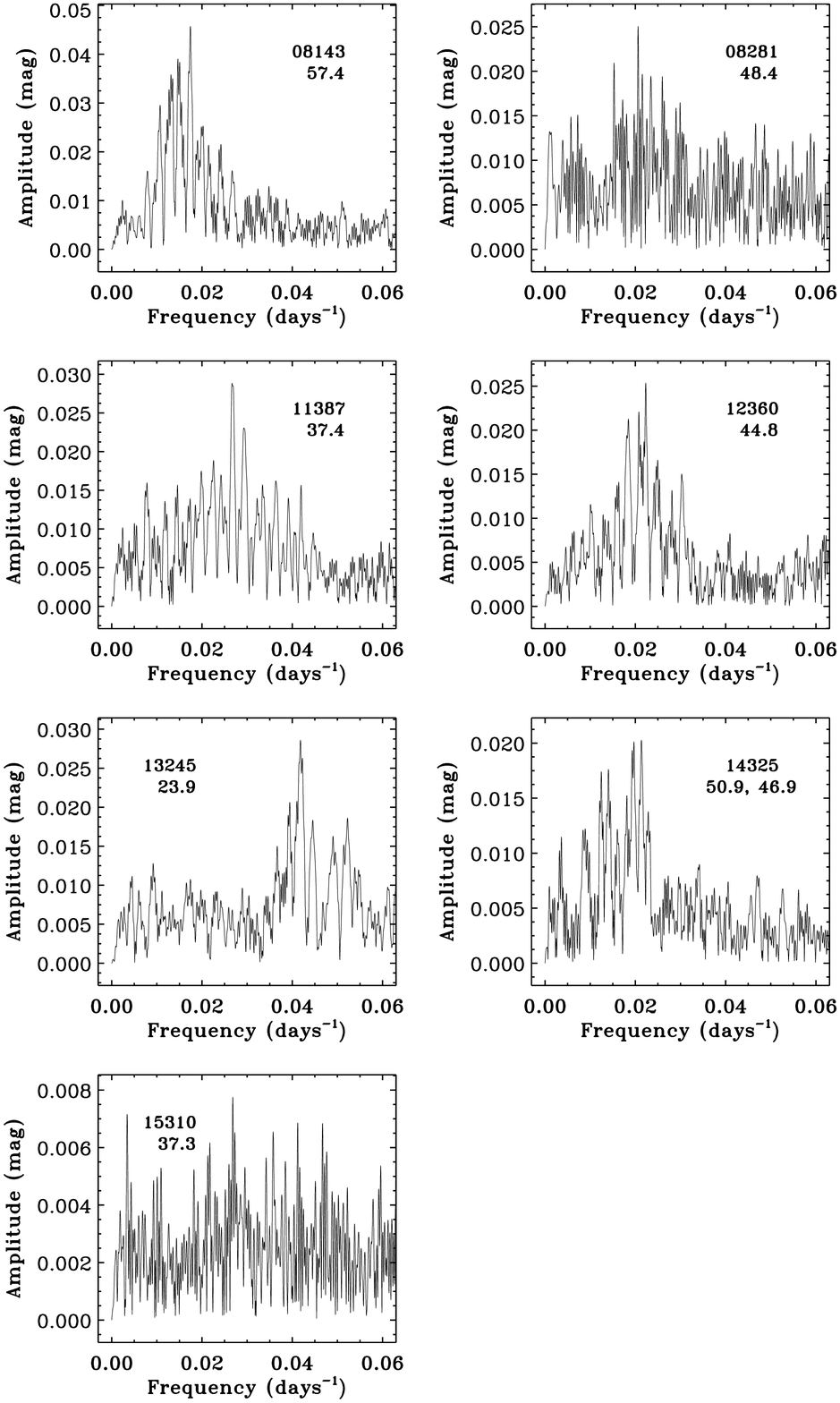}
\caption{ The frequency spectrum of the seasonally-normalized observations, phased with their dominant periods, for the seven objects with reasonably well-established periods.  The period is listed under the abbreviated name of each, in units of days.  
\label{freqspec}}
\epsscale{1.0}
\end{figure}

\begin{figure}\figurenum{3}\epsscale{1.7} 
\plotone{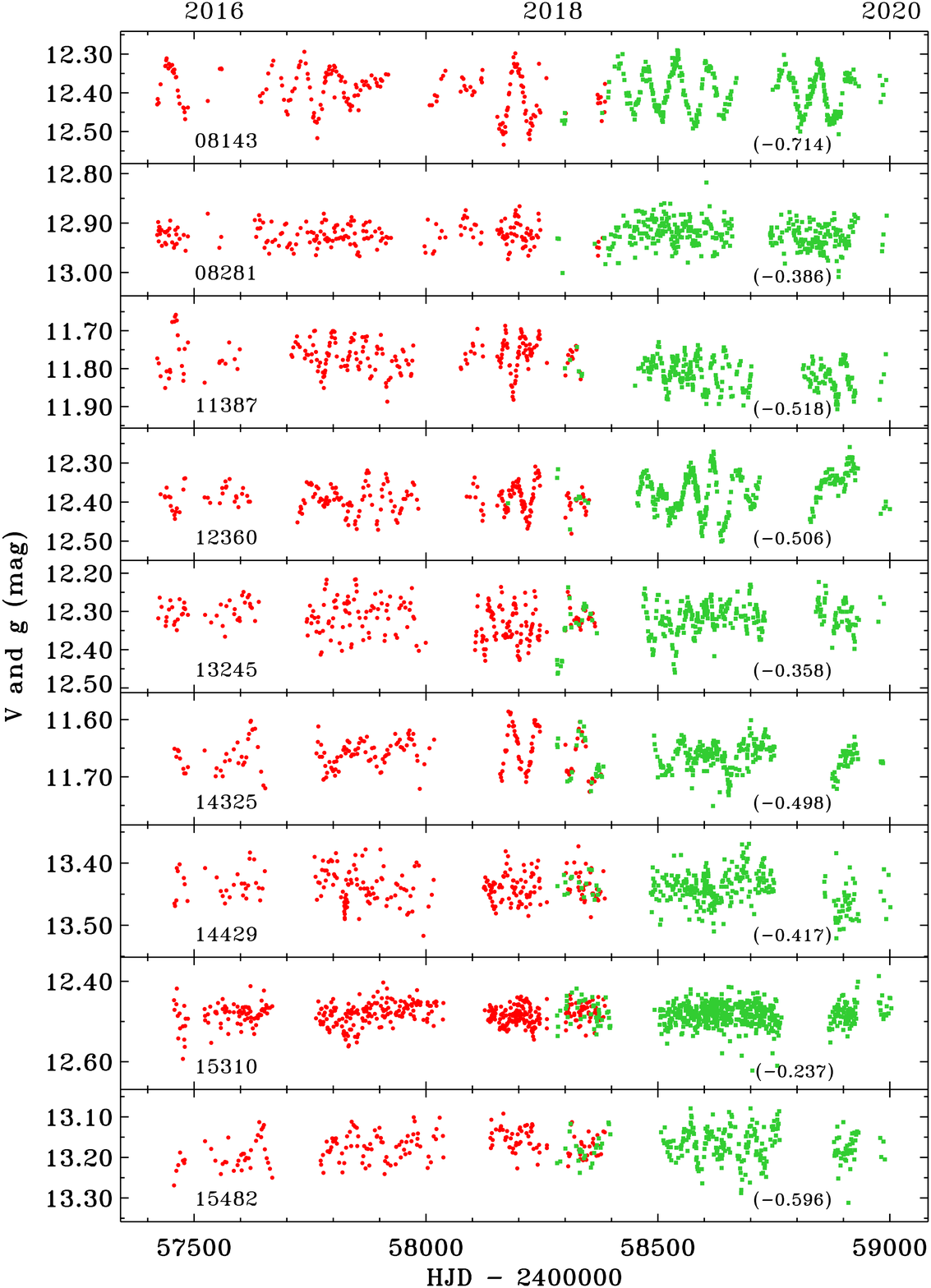}
\caption{The ASAS-SN {\it V} light curves (red circles) from 2015 to 2018 and {\it g} light curves (green squares) from 2018 to 2020, with the later offset to the level of the {\it V} light curve for the interval of overlap (2458280$-$2458400).  The amount of the offset in units of mag is listed in parentheses. 
Note that the light curves of IRAS 14429$-$4539, 15310$-$6149, and 15482$-$5741 are each contaminated by a nearby star in the large ASAS-SN photometric aperture.  
\label{fig3}}
\epsscale{1.0}
\end{figure}

The analyses of the light curves are presented below.  
We give a summary of the investigation of each and the main results.  
The periodogram of each of the stars for which a period has been determined are shown in Figure~\ref{freqspec}.

\placefigure{freqspec}

\subsection{Individual Objects}

{\it IRAS 08143$-$4406} $-$ 
Visual inspection of the SARA light curves reveals clear cyclical variations, which can be seen in Figure~\ref{fig1}.
The analyses of the various light curves of IRAS 08143$-$4406 show cyclical, periodic variations with similar periods.
Periodogram analyses of the SARA {\it V} and {\it R}$_C$ data from 2010-2018 yielded consistent, significant periods of {\it P}$_1$ = 73.6$\pm$0.1, {\it P}$_2$ = 58.9$\pm$0.1, and {\it P}$_3$ = 68.4$\pm$0.1 days in each, with {\it P}$_1$ being the period with the largest photometric amplitude and highest S/N.
A similar period of 73.6$\pm$0.2 days was also found in the {\it V$-$R}$_C$ color data.
The much higher density ASAS-SN data yielded periods of 57.4$\pm$0.1 and 76.1$\pm$0.2 days in {\it V} and 54.6$\pm$0.2 and 61.3$\pm$0.2 days in {\it g}, over shorter time intervals of two to three seasons.  An analysis of the combined ASAS-SN {\it V} and {\it g} data over five seasons, with an empirically-determined color offset, yielded periods 73.3$\pm$0.1, 58.3$\pm$0.1, and 66.4$\pm$0.1 days, very similar to the values found in the SARA light curves.
When we combined the {\it V} data from the SARA light curve with the ASAS-SN data and analyzed them, we determined values identical to the first two periods of the ASAS-SN {\it V} data alone, which perhaps is not surprising, since the ASAS-SN data are much more numerous.  This is the periodogram plot shown in Figure~\ref{freqspec}.

\begin{figure}\figurenum{5}\epsscale{2.0} 
\plotone{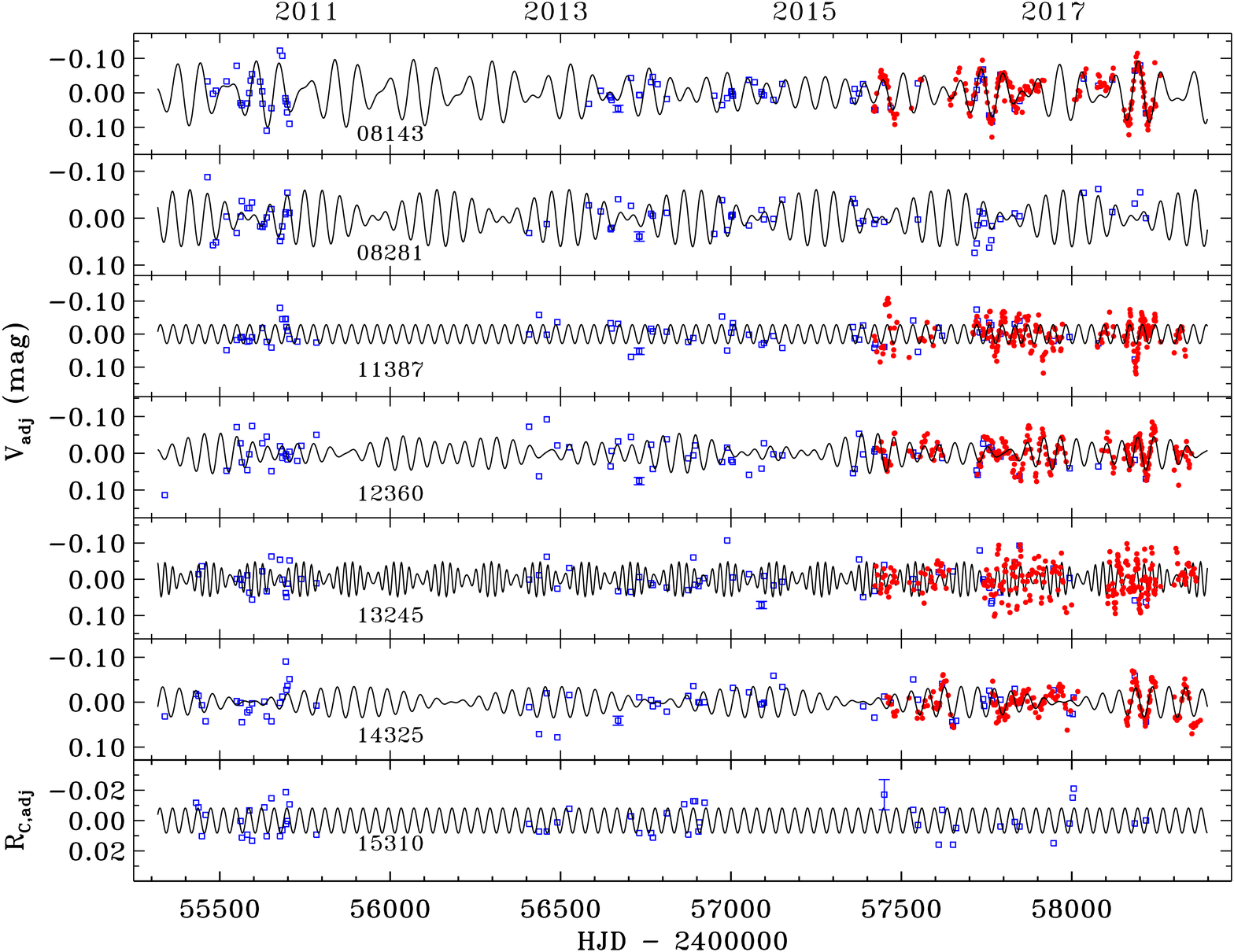}
\caption{The combined SARA (blue squares) and ASAS-SN (red circles) light curves, normalized as described in the text, and fitted by the periods, amplitudes, and phases recorded in Table~\ref{periods}.  They are all {\it V} light curves except for IRAS 15310$-$6149, which is {\it R}$_C$.  Also shown for each is an example of a typical error bar, located near the center of the light curve.  
\label{LC_fits}}
\epsscale{1.0}
\end{figure}

\begin{figure}\figurenum{6}\epsscale{1.2} 
\plotone{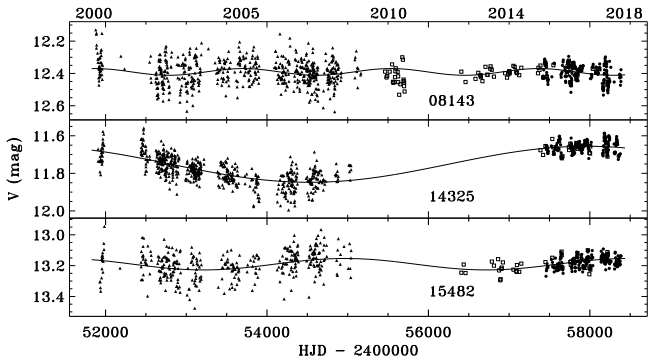}
\caption{The combined ASAS-3 (triangles), SARA (squares), and ASAS-SN (circles) {\it V} light curves from 2010 to 2018 for the three objects with multi-year, periodic variability. 
The light curves are fitted by the periods, amplitudes, and phases recorded in Table~\ref{periods}.  
The average uncertainties for the ASAS-3 data range from $\pm$0.04$-$0.05 mag and for the SARA and ASAS-SN data they range from $\pm$0.010$-$0.015 mag.
\label{long-P}}
\epsscale{1.0}
\end{figure}

There is an indication of a long-term variation seen in the ASAS-3 light curve.  
An analysis of a combined ASAS-3, SARA, and ASAS-SN {\it V} light curve from 2000 to 2018 yielded a significant period of 5.0$\pm$0.1 years (with the individual SARA and ASAS-SN data points given three times the weight of the much less precise ASAS-3 data points).
We conclude that there are at least two pulsation periods in the data sets, 73 and 58 days, with a period ratio ({\it P}$_2$/{\it P}$_1$=0.79), and a long-term period of 5.0 yrs. 
The period analysis results of the combined, seasonally-normalized SARA and ASAS-SN {\it V} light curve, together with associated amplitudes and phases, are listed in Table~\ref{periods}.
In Figure~\ref{LC_fits} is shown the model fit to the combined, seasonally-normalized SARA and ASAS-SN light curve, neglecting three SARA seasons with only a few observations that one could not reliably normalize.  
This produces a reasonably good fit to the light curve, with an average standard deviation of 0.026 mag (see last column in Table~\ref{periods}).
The fit to the long-term period of 5.0 years is shown in Figure~\ref{long-P}, and it gives a good fit to the long-term variations.
The amplitude for this fit is small, 0.02 mag, and the parameters are also listed in Table~\ref{periods}.

\placetable{periods} 

\placefigure{LC_fits}

\placefigure{long-P}

{\it IRAS 08281$-$4850} $-$  
Analyses of the camera-normalized SARA light curves yield periods of 48.4$\pm$0.1 and 55.6$\pm$0.1 days in both the {\it V} and {\it R}$_C$ light curves.
Both the ASAS-SN and the ASAS-3 light curves are contaminated by a nearby star that contributes over 60\% of the light. This greatly dilutes the light variability and makes the results derived from the study of those light curves much less reliable.  Thus, for IRAS 08281$-$4850, we base our period study solely upon the results of the analysis of the SARA {\it V} and {\it R}$_C$ data.  
The {\it V} light curve, fitted with these two periods and their associated amplitudes and phases, is shown in Figure~\ref{LC_fits}. 
These periods produce a reasonably good fit to the light curves, although some of the amplitudes are off, with an average standard deviation of 0.020 mag.
The period ratio is 1.15.

{\it IRAS 11387$-$6113} $-$
No significant period is found in the SARA data.
However, periodicity is found in the high density ASAS-SN data, 37.4$\pm$0.1 days in the {\it V} light curve over three seasons and 28.5$\pm$0.1 days in the {\it g} light curve over approximately two seasons.  
Analyzing the two data sets, combined together with an empirically-determined offset, results in two significant periods similar to those found individually, 37.0$\pm$0.1 and 27.6$\pm$0.1 days.
Combining the SARA and ASAS-SN {\it V} data and analyzing this combined light curve results in a period of 37.4$\pm$0.1, the same as found for the much more numerous ASAS-SN data when analyzed alone.  We will tentatively claim this to be the pulsation period of the object, but this needs further confirmation.
In Figure~\ref{LC_fits} is shown the fit to this combined {\it V} light curve. 
The period appears to fit the light curve reasonably well, but the fit to the changing amplitudes is not good.

{\it IRAS 12360$-$5740} $-$
Neither of the SARA {\it V} or {\it R}$_C$ light curves were found to be periodic.  However, the high density ASAS-SN light curves show clear cyclical variations over most seasons, although with varying amplitudes, as can be seen in Figure~\ref{fig3}.  
Analysis of the ASAS-SN {\it V} light curve, covering two and a half seasons at high density, reveal two periods of similar strengths, 44.4$\pm$0.1 and 54.2$\pm$0.2 days.  These give a reasonably good fit to the light curve.  The ASAS-SN {\it g} light curve covers about two seasons and results in a period of 49.4$\pm$0.1 days, which lies partway between the two {\it V} periods.  
Combining the ASAS-SN {\it V} and {\it g} data, with an offset to the {\it g} data, results in a single period of 47.6$\pm$0.1 days.
An analysis of the combined SARA and ASAS-SN {\it V} light curves, covering 2010 to 2018 with the heavy emphasis on the 2016$-$2018 data, resulted in a dominant period of 44.8$\pm$0.1 days, a secondary period of 48.2$\pm$0.1 days, and a barely significant third period of 53.6$\pm$0.1 days (S/N=4.0). 
Assessing the results of these various analyses of the light curves of IRAS 12360$-$5740, in particular the combined SARA and ASAS-SN {\it V} light curve, 
we conclude that there is a dominant period of {\it P}$_1$ = 44.8$\pm$0.1 days, a secondary period of {\it P}$_2$ = 48.2$\pm$0.1 days, and evidence for a third period of {\it P}$_3$ = 53.6$\pm$0.1 days.  These result in a period ratio ({\it P}$_2$/{\it P}$_1$) of 1.08. 
These periods give a relatively good fit to the combined, normalized (adjusted) SARA and ASAS-SN {\it V} light curves, as shown in Figure~\ref{LC_fits}.

{\it IRAS 13245$-$5036} $-$ 
The results of the analysis of the light curves of this object are not so clear.  The analysis of the SARA {\it V} and {\it R}$_C$ light curves revealed a period of 10.7 $\pm$0.1 days in each.  However, the higher density ASAS-SN {\it V} data, which cover three full seasons, yielded a period of 23.8$\pm$0.1 days and a barely significant second period of 20.5$\pm$0.1 days (S/N=4.0) and the {\it g} data, which cover one full season and smaller parts of two others resulted in a period of 48.9 $\pm$0.2 days, about twice that of the {\it V} light curve.  
Analyzing the combined ASAS-SN {\it V} and {\it g} light curves did not result in a significant period, 
while the analysis of the combined SARA and the ASAS-SN {\it V} light curves resulted in the two periods found for the ASAS-SN data above, but with the second period no longer significant.  
Assessing these results, we suggest a likely period of 23.8 days, but this surely needs to be confirmed by further observations. 
The light curve fit to the combined SARA and ASAS-SN {\it V} light curve alone is shown in Figure~\ref{LC_fits}.  The fit is not good to the SARA data.

{\it IRAS 14325$-$6428} $-$
The results of the analysis of the light curves of IRAS 14325$-$6428 are also not clear or consistent.  
The SARA data show some relatively large differences in brightness between data obtained with the three different cameras.  This may be the result of real variations with time or it may be a calibration effect in the data due to the redness of the comparison stars used for this object.   
No periodicity is found in the SARA light curves, even when we adjusted the data from the three cameras to the same brightness level.  
The ASAS-SN {\it V} light curve does show a clear cyclical behavior, especially in the 2017$-$2018 season, which possesses the largest amplitude (see Figure~\ref{fig3}).  
The periodogram analysis results in a primary period of 51.4$\pm$0.1 days and secondary period of 46.7$\pm$0.1 days, and these produce a good fit to the adjusted {\it V} light curve.
The ASAS-SN {\it g} light curve also shows a cyclical pattern in the partial season of overlap, 2017$-$2018, but does not for the rest of the data, and no significant period is found in the {\it g} light curve.  
Analysis of the combined SARA and the ASAS-SN {\it V} light curve results in two period of equal strength, 50.9$\pm$0.1 and 46.9$\pm$0.1 days, similar to what was found in the ASAS-SN light curve alone.   
Thus, while there is clear evidence of periodicity in the data in some years, this is not well maintained throughout the interval of observations.  
Shown in Figure~\ref{LC_fits} is the fit to the normalized SARA and ASAS-SN {\it V} combined light curve.  The fit to the periodicity appears to be reasonably good, 
but the fit is not very good to the early SARA data, especially the amplitudes. 

Visual inspection of the ASAS-3 {\it V} light curve shows that it decreases in brightness by $\sim$0.17 mag, with an indication of a sinusoidal shape.  
Combining this with the SARA and ASAS-SN {\it V} light curve data, and giving these more precise observations a weight equal to three times that of a ASAS-3 observation,   results in a long period of 19 years.
However, the SARA data from the first two cameras systematically deviate from the light curve fit.  As noted above, there are reasons to suspect some possible systematic effects, 
so neglecting these data we find a long-term period of 18.8$\pm$0.2 yrs in the {\it V} light curve. 
This is shown in Figure~\ref{long-P}, and the long-term variations are well fitted with this sine curve.  
However, caution is required in accepting this period, since the value is approximately equal to the observing interval.  A longer observing interval is need to confirm this period.

{\it IRAS 14429$-$4539} $-$
The various SARA and ASAS-SN data sets were analyzed separately and in various combinations, but not significant periods of pulsation were found for this star.  The variations in brightness of the star during a season reached a maximum value of 0.12 mag in the ASAS-SN {\it V} data of 2018$-$2019.
The ASAS-3 data are $\sim$0.3 mag brighter than the ASAS-SN and SARA {\it V} data, which might suggest some long term variability in the star or perhaps contamination by the brightness of another star in the aperture.

{\it IRAS 15310$-$6149} $-$
Examination of the SARA {\it V} and {\it R}$_C$ light curves shows no obvious indication of cyclical behavior within a season. 
However, there are significant differences in the brightness levels between the different cameras, with the observations made with camera 2 in the 2013-2014 and 2014-2015 seasons being 0.025 mag brighter.  While this may indicate a real variation in brightness over the eight year observing interval, it might also indicate a systematic offset in the differential brightness of the star as measured with the different cameras.  .
Unfortunately, we are not able to use the ASAS-SN data to help distinguish between these to possibilities for reasons indicated below.
We proceeded in the analysis by normalizing the SARA data for each camera to the same brightness level.
These light curves cover a small range in brightness, with a maximum in any camera reaching 0.046 mag in {\it V} and 0.037 mag in {\it R}$_C$.  Analyses of both light curves, normalized by camera, suggest a period of 37.3$\pm$0.1 day, with the period being significant in {\it R}$_C$ (S/N=4.2) but not in {\it V} (S/N=3.7).  The fits to the light curve are fair, but not good, with small amplitudes, 0.009 mag in {\it V} and 0.008 mag in {\it R}$_C$.  
Thus we regard this as a tentative period.
The {\it R}$_C$  periodogram is shown in Figure~\ref{freqspec}.
No significant periodic variations are found in the ASAS-SN or ASAS-3 light curves.  However, they are each severely contaminated by a nearby brighter star 
{\it V}=12.69 located 11$\arcsec$ east of the program star.  The brightnesses recorded for the program star in these two {\it V} light curves are 12.4$-$12.5 mag, compared with our SARA measurements of {\it V} = 13.6 mag with a smaller aperture that excludes the nearby star.
Thus 65$\%$ of the light in the ASAS-SN and ASAS-3 measurements originates in the contaminating star. 
Shown in Figure~\ref{LC_fits} is the fit to the adjusted SARA {\it R}$_C$ light curve.

{\it IRAS 15482$-$5741} $-$ 
The SARA observations began in 2013 and utilized only cameras 2 and 3, and thus there are not a lot of observations in each filter (39).  The object appears to increase in brightness by 0.045 mag during the parts of five seasons observed.  Period analyses revealed the same dominant period in each of the {\it V} and {\it R}$_C$ light curves, 61.5$\pm$0.2 
and the same almost significant secondary period of 41.1$\pm$0.2 mag.  These give a reasonably good fit to the normalized light curves.  
Both the ASAS-SN and ASAS-3 light curves have been significantly contaminated by nearby field stars.
The ASAS-SN {\it V} light curve is 0.3 mag brighter than the SARA {\it V} light curve during the interval of overlap (2016$-$2018) and the ASAS-3 light curve is 0.7 mag brighter.  These correspond to contamination of light at the 25 $\%$ and 50 $\%$ levels, respectively.
In spite of the contamination, they were still useful.  
The ASAS-SN {\it V} light curve shows no evidence of periodicity in the light variations. 
In contrast, the {\it g} light curve can be seen visually to vary cyclically during the 2018$-$2019 season, and a period of 45.0$\pm$0.2 days was found in the entire {\it g} light curve.
Thus, no consistent pulsation period is found among the various data sets, with values of 61.5 days (SARA, 2013$-$2018), 45.0 days (ASAS-SN {\it g}, 2018$-$2020), and no significant period found in the ASAS-SN {\it V} (2016-2018) light curve.  Therefore, we withhold judgement at present on the value of periodicity in the short-term variability of this object.  

The ASAS-SN {\it V} light curve does serve to confirm the increase in brightness seen in the SARA light curves, 
and  the increase in brightness seen in these two light curves seems to be part of a long period cyclical behavior in light.  
This is seen in the combined SARA and ASAS-SN light curves from 2013$-$2018 and also in the ASAS-3 light curve from 2000$-$2009.  Combining these three {\it V} light curves, with offsets in their brightness level to adjust for the contamination mentioned, we determine a long period light variation of 9.6$\pm$0.2 yrs,  
again giving the SARA and ASAS-SN data three times the weight of the less precise ASAS-3 data.
This is shown in Figure~\ref{long-P}.  The data are well fitted with an amplitude of 0.04 mag.

\section{DISCUSSION}

\subsection{Spectral Energy Distributions}
\label{seds}

The spectral energy distributions (SEDs) observed for all nine program objects show the double peaks characteristic of PPNe, although for IRAS 14429$-$4539 they are less distinct.  One peak occurs at $\sim$1 $\mu$m and arises from the photosphere of the star, subsequently reddened by circumstellar and interstellar dust.  The other peak occurs at $\sim$25 $\mu$m and is due to re-radiation from the circumstellar dust.
These are shown in Figure~\ref{SEDs}.  
In all nine cases, the observed peak in the mid-infrared is higher, an indication of the large effect of the dust.

\placefigure{SEDs}

These SEDs can used together with parallax measurements from the {\it Gaia} spacecraft to determine approximate luminosities for the objects.  
One must, of course, take into account the interstellar extinction.  The effect of extinction is expected to be quite large, as these objects are located near the galactic plane, 
with seven of the objects having $ \left|{\it b}\right| $ $<$ 6$\arcdeg$ and the other two having {\it b} $\sim$ 12$\arcdeg$.
While we are not able to determine directly the value of the interstellar extinction in the {\it V} bandpass, {\it A}$_V$, we can find useful bounds on these values for each object from two published studies.  
\citet{cap17} produced 3D reddening maps of the galaxy reaching out to distances of several parsecs.  For the seven objects that are near the plane of the galaxy, the maps reach to distances of 2.0 to 2.8 kpc and for the other two they reach to 1.4 kpc.  These are much less than the Gaia distances found for these objects and serve to set a lower limit to the interstellar extinction.
From the study of \citet{sf11}, which is based on background galaxies, upper limits on the interstellar {\it A}$_V$ were determined. 
The values for the distance and extinction for each object are listed in Table~\ref{sed_table}.  

\placetable{sed_table}
  
The observed SEDs were then dereddened using each of these values of the interstellar extinction and assuming a reddening value ({\it A}$_V$/{\it E}({\it B$-$V})) of 3.1.
The reddening function of \citet{rie85} was used and extended to longer and shorter wavelengths.  
By integrating these dereddened SEDs, we determined upper and lower limits for the luminosities of the objects.  
Note that we do not correct explicitly for the effect of the circumstellar dust, since the radiation that is absorbed is re-radiated in the mid-infrared back into our line of site.
In this procedure, we make the assumption that the dust is distributed isotropically around the star.
If this is not the case, then the visible light contribution may be somewhat larger or smaller than estimated.  
We also make the reasonable assumption that the dust is optically-thin in the mid-infrared.   
In Figure~\ref{SEDs} is also shown the SED values dereddened for interstellar extinction using the upper limit values of \citet{sf11}.  
In four of the cases the upper limit for the extinction has a value of A$_V$ $\ge$ 2.0, and for those stars the dereddened visual observations are much brighter and become comparable to or even brighter than the mid-infrared observations.
We have fitted all but one of these with two blackbody curves, representing the emission from the dereddened photosphere (corrected for interstellar but not circumstellar extinction) and from the cool, circumstellar dust.  The dust temperatures range from 140$-$250 K.  The photospheric temperatures range from 4800$-$9500 K, lower than expected for their spectral type (see Tables~\ref{object_list}, \ref{results}), but consistent with the fact that they are not corrected for circumstellar reddening.  
The SED for IRAS 14429$-$4539 is not as distinctively bimodal as the others, and two blackbody curves do not produce a good fit.  We have instead fitted the visible and near-infrared curves separately, in addition to the mid-infrared.  This gives a better fit, although even then there is excess flux in the 3$-$8 $\mu$m region.  The situation is improved if we imagine that there is a large amount of circumstellar extinction (A$_V$$\approx$2 mag, instead of 0.4 mag), and then a fit can be made to the visible and near-infrared flux with a single photospheric temperature, similar to what is seen for IRAS 08281$-$4850, which has a somewhat similar shape to its SED.  However, even with the assumption of a large amount of circumstellar extinction, there would still be excess flux in the 3$-$8 $\mu$m region, suggesting warm dust.
A similar excess in this wavelength region is seen in the SED of IRAS 15310$-$6149, also suggesting warm dust.

Gaia distance determinations are available for eight of the nine objets \citep[{\it Gaia} Early Data Release (EDR3);][]{bai21} and range from 3.8 to 14.2 kpc, with uncertainties generally ranging from 10$-$20$\%$.\footnote{We used the \citet{bai21} distances derived from geometric priors, since those do not include assumptions about the color and apparent magnitude of these post-AGB stars, as do the photometric priors.  The earlier
{\it Gaia} Data Release 2 \citep[DR2;][]{bai18} distance values not only have larger uncertainties, 20$-$30$\%$, but in many cases are very significantly different, with the DR2 distances ranging from 0.5 to 1.6 times the EDR3 distances.}  
The distances and resulting luminosities are listed in Table~\ref{sed_table}.  If we assume that the actually luminosities are approximately half way between these two extinction limits, then we find that they range from 600 to 10,000 {\it L}$_{\sun}$.
However, the upper limit for IRAS 11387$-$6113 is very uncertain due to the uncertainty in its interstellar extinction, and it luminosity might be even higher.
As noted above, the uncertainty in the geometry of the circumstellar dust will have some effect on the uncertainty in the calculated luminosity.  
The listed uncertainties in the distances generally range from about 10$\%$ lower to about 20$\%$ higher than the  best-determined values.  These lead to an uncertainty in the luminosity of approximately 0.8 to 1.4 times the determined value.

Two of the objects have rather low luminosity values for PPNe, IRAS 08143$-$4406 (2100 {\it L}$_{\sun}$) and 15482$-$5741 (600 {\it L}$_{\sun}$).  
Stellar models by \citet{milb16} result in post-AGB luminosities of 2500 to 25,000 {\it L}$_{\sun}$ for core masses of 0.53 to 0.83 M$_{\sun}$, respectively. 
Given the uncertainties in the luminosities, IRAS 08143$-$4406 may simply represent a PPN whose distance is underestimated or whose extinction is near the upper limit.  
IRAS 15482$-$5741, on the other hand, has too low a luminosity be be a post-AGB object.  We will return to discuss this star later.
IRAS 11387$-$6113 also deserves special mention.  The object is directly in the plane of the galaxy ({\it b} = 0$\arcdeg$) at a distance of 5.0 kpc.  The upper limit of the extinction in this direction is {\it A}$_V$ = 7.0 mag, a value much larger than that of any of the other program objects.  
Correcting the SED for this large value of interstellar extinction yields a luminosity of 130,000 {\it L}$_{\sun}$, far larger than that of a post-AGB star and rather one appropriate for a very luminous supergiant.  
If we assume a smaller value of {\it A}$_V$ = 3.0 mag, we arrive at a value of {\it L} = 6,600 {\it L}$_{\sun}$.  
Thus seven of the eight objects with distance measurements can be considered to be PPNe based on the shape of the SEDs and on their luminosities, to within the uncertainties of the measurements.  Future improvements in the distance measurements and, in one case, in the interstellar extinction value, will make these evolutionary classifications more secure.

\subsection{Oxygen- or Carbon-rich Chemistries?}

PPNe have been classified as either oxygen-rich or carbon-rich, based on which of the two is dominant in the stellar photosphere and the circumstellar envelope.  
In a few rare cases, there is a mixed chemistry, with both oxygen- and carbon-rich features present (e.g., IRAS 08005$-$2356 $-$ \citet{bak97}; IRAS 19306$+$1407, 20462$+$3416, 22023$+$5249 $-$ \citet{cer09}).
Atmospheric abundance studies based on high-resolution spectra have revealed the carbon-rich photospheres of IRAS 08143$-$4406, 08281$-$4850, 13245$-$5036, 14325$-$6428, and 14429$-$4539 \citep[and references therein]{des16} and the oxygen-rich photospheres of IRAS 12360$-$5740 and 15482$-$5741 \citep{per12}.

Several of the objects also have mid-infrared spectra which reveal the chemistry of the circumstellar dust.  
The carbon-rich nature of the circumstellar dust is evidenced by the PAH features seen in the IRAS 14429$-$4539 and 15482$-$5741 and supported by the 21 and 30 $\mu$m features seen in both of these and also in IRAS 13245$-$5036 \citep{cer11}.  
IRAS 11387$-$6113 has strong silicate features, on the basis of which we can classify it as O-rich, but it also appears to have some weak PAH features \citep{ram17}, suggesting a mixed chemistry. IRAS 14325$-$6428 has a strong and wide 11 $\mu$m feature, the origin of which is uncertain \citep{ram17}.

We have not found a definitive indicator of the chemistry of IRAS 15310$-$6149, but we see a suggestion of C$_2$ bands in its optical spectrum \citep{suarez06}, suggesting that it is C-rich, but this should be confirmed.
IRAS 15482$-$5741 presents the anomalous situation in which the photosphere is O-rich and the circumstellar envelope is C-rich.  
The circumstellar envelope of IRAS 14429$-$4539 has been detected in the gaseous phase in CO \citep{nym92} but not OH \citep{hu94}, consistent with its C-rich chemistry.

\subsection{Pulsational Variability}

PPNe vary in brightness, with periods ranging from $\sim$35 for early-F to 160 days for late G spectral type stars. 
These variations have been shown to be due to pulsations \citep{hri13,hri18}.
In this study, eight of the nine objects show evidence of periodic variability due to pulsations, although for one (IRAS 15482$-$5741) the period is highly uncertain since its value was not consistent among the different data sets. 
For three of the objects, the ones with the shortest periods, the values are regarded as provisional.  
It was also the case in a previous study that the two short period objects with {\it P} $\approx$35-37 days had the most uncertain  periods  \citep[IRAS 07134+1005, 19500$-$1907;][]{hri10}.  
The uncertainty in the periods of these will hopefully be resolved by several additional seasons of ASAS-SN data.
This is the first documented evidence of periodic variability for seven of these objects.

\begin{figure}\figurenum{8}\epsscale{0.8} 
\plotone{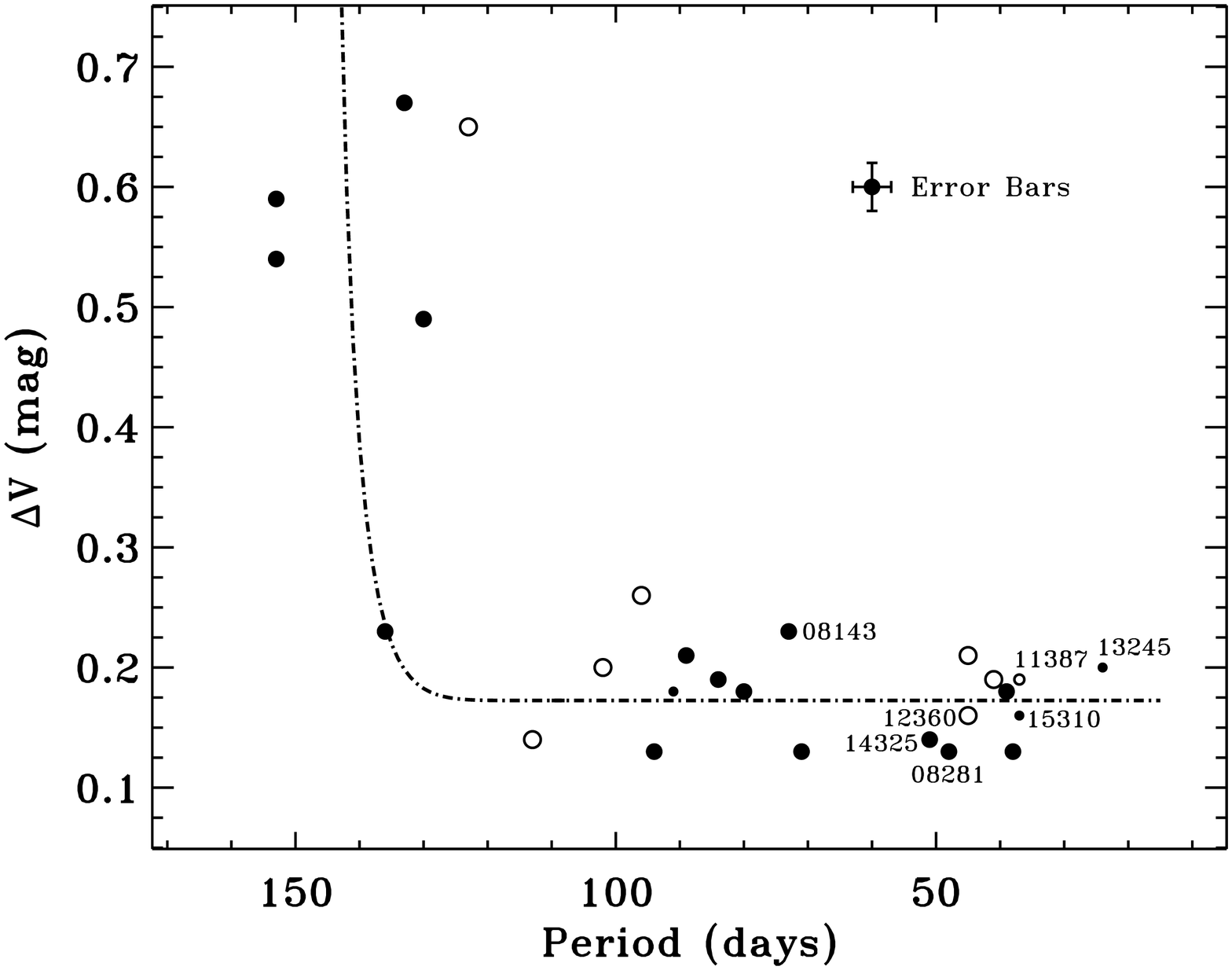}
\plotone{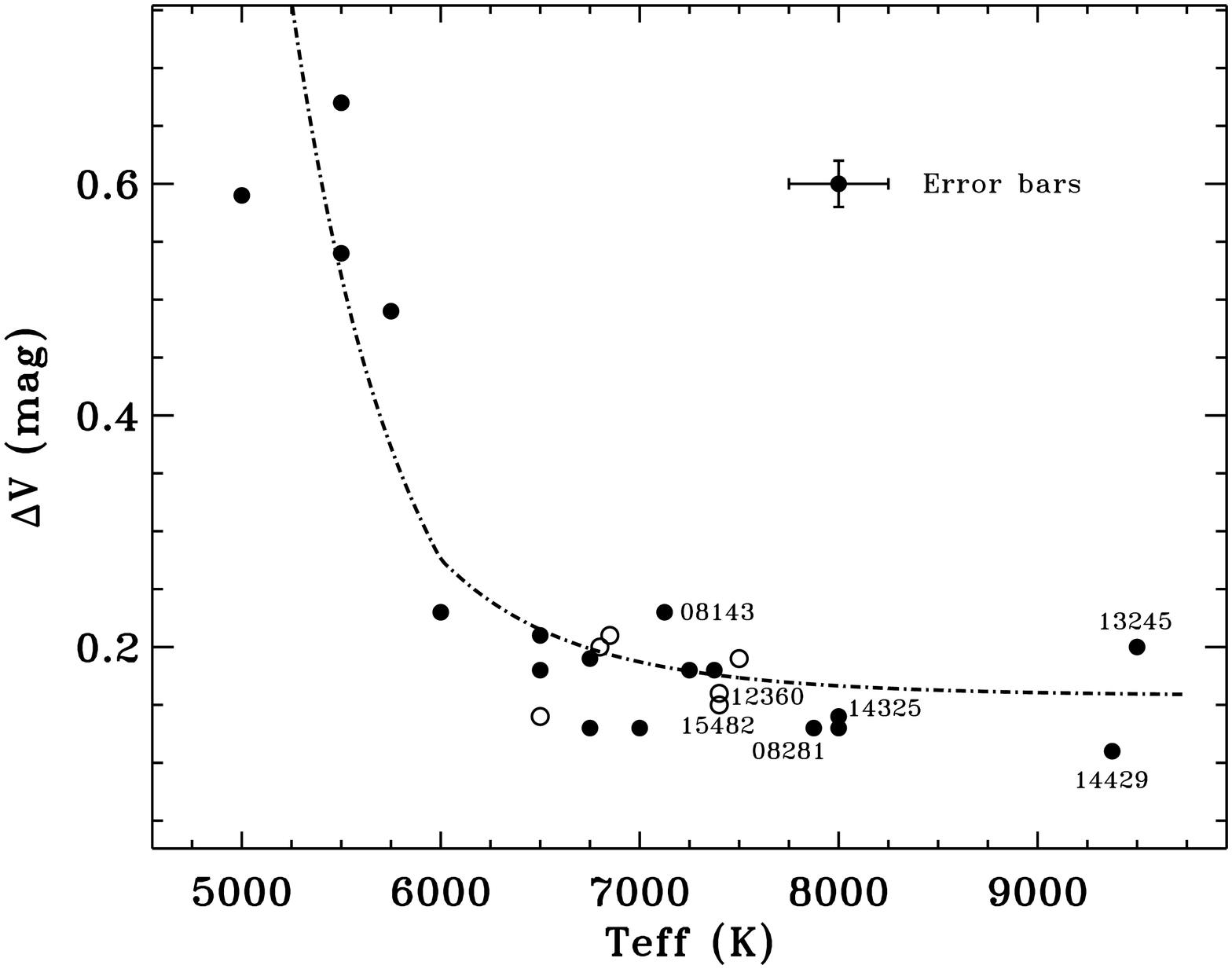}
\plotone{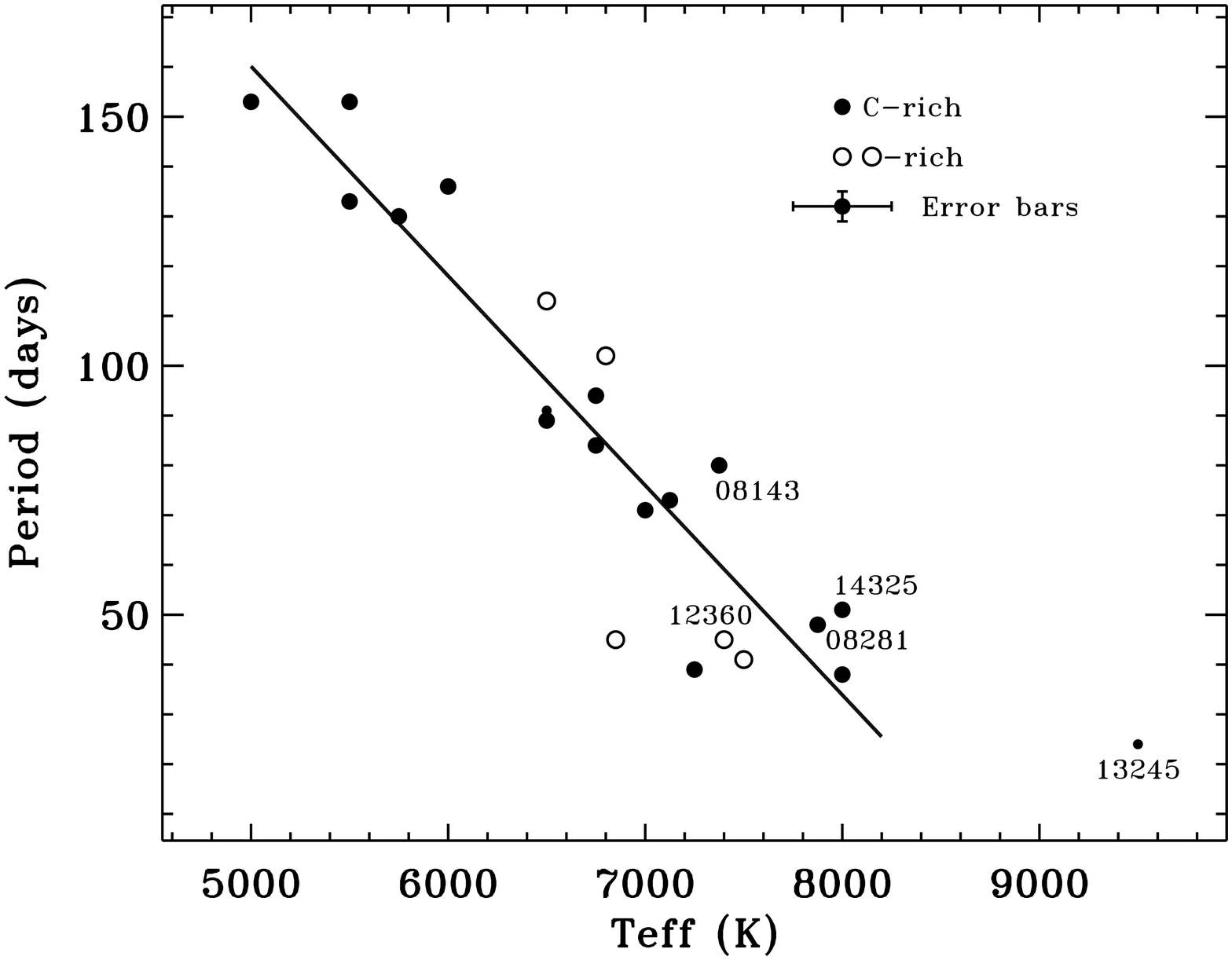}
\caption{Plots comparing the pulsation properties (period and maximum amplitude in a season) and temperatures of these new program sources, along with those of other PPNe.
The dashed lines are free-hand representations of the trends in the data.  The solid line in the bottom panel is a least squares fit to the data, excluding the one point at higher temperature.  The new results from the stars in this study are labeled.  Those in which the periods are less certain are shown with smaller symbols.  Typical error bars are shown in each panel, with an uncertainty of $\pm$3 days, $\pm$0.02 mag, and $\pm$250 K. 
\label{PVT_plots}}
\epsscale{1.0}
\end{figure}

\begin{figure}\rotate\figurenum{7}\epsscale{2.2} 
\plotone{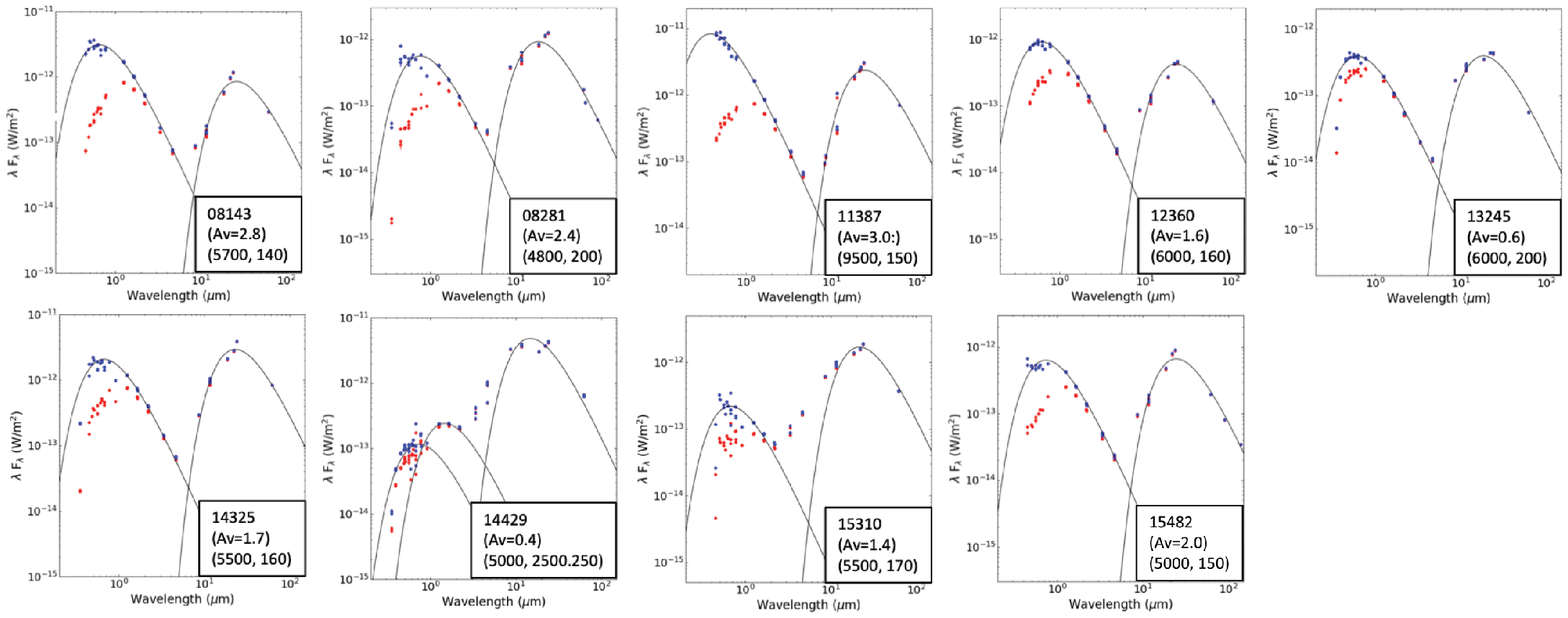}
\caption{The spectral energy distributions of the program objects. 
The observed flux densities are plotted in red and the values de-reddened for interstellar extinction are plotted in blue.  The visual interstellar extinction values are listed, using the upper limits.
Two black body curves are shown for each object, one in the visible/near-infrared representing a fit to the photospheric emission de-reddened for interstellar (but not circumstellar) extinction, and the other, in the mid-infrared, representing the fit to the circumstellar dust emission. The temperatures of these are also listed in the panels.  
Plotted are the {\it IRAS}, {\it AKARI}, {\it WISE}, {\it 2MASS}, Johnson {\it B,V,J,H,K}, Sloan Digital Sky Survey, {\it Gaia}, and SkyMapper \citep{wolf18} data, with error bars, as listed in SIMBAD-VizieR.
\label{SEDs}}
\epsscale{1.0}
\end{figure}

For the seven more-certain periodic objects, the values range from 24 to 73 days, and several are among the shortest periods measured thus far for PPNe.
The variations in brightness are not large, 
with maximum seasonal variations ranging from 0.11 to 0.23 mag ({\it V}).
These are consistent with the low-level variations seen in other PPNe with periods less that 120 days, and are much smaller than the variations seen in most PPNe of longer periods, 
which can be as large as 0.5$-$0.7 mag in {\it V} \citep{hri10}.
The period and maximum seasonal brightness variation of each are listed in Table~\ref{results}.
In Figure~\ref{PVT_plots}a we show these results graphically.

\placetable{results}

Of even greater interest is the comparison of the period and maximum brightness variation with spectral type and photospheric temperature.  These properties are also listed in Table~\ref{results}.
Comparing their maximum brightness variation with their temperatures, one sees that their small variations in brightness are similar to those of other PPNe with {\it T}$_{\rm eff}$ $>$ 6000~K (Fig.~\ref{PVT_plots}b).
In comparing their spectral types with period, one sees that those of A spectral types have the shortest periods and those of mid- and late-F spectral types are longer.  This trend of shorter periods with hotter stars has been seen our previous studies of PPNe. 
Five of the seven periodic PPNe have photospheric temperature ({\it T}$_{\rm eff}$) derived from atmospheric model fits to high-resolution, visible-band spectra.  They show a trend of shorter period with higher temperature.  This is seen in  Figure~\ref{PVT_plots}c, where we have again included the values from other periodic PPNe.
These new objects fit with the linear trend for objects with {\it T}$_{\rm eff}$ $<$ 8000~K.
However, they obviously could not continue to do so to much higher temperatures, and we see here the change in slope for the object with {\it T}$_{\rm eff}$ = 9500~K.  
Note that in these plots we have distinguished the stars according to their chemistry as C-rich or O-rich.

\placefigure{PVT_plots}

The four PPNe in this study with well-determined periods are all found to have multiple periods, with period ratios, {\it P}$_2$/{\it P}$_1$, of 0.79$-$1.15.  This is common among PPNe, and previous studies have found typical period ratios, {\it P}$_2$/{\it P}$_1$, ranged from 0.8 to 1.2 \citep[e.g.,][]{hri13,hri15}.  

As the stars pulsate and change in brightness, we find that most of them show a small systematic change in color, becoming slightly redder when fainter.  Comparing the color change, $\Delta$({\it V$-$R}$_C$), with the temperature of the star, using the color$-$temperature relationship for supergiants \citep{cox00}, indicates small temperature changes of 350$-$500 K.

\subsection{Comparison with Pulsation Models}

Only a few hydrodynamic models of stellar pulsation in post-GB stars are available.  
\citet{fok01} studied non-linear, radiative pulsation models, focusing on a narrow range of cool temperatures (5600$-$6300 K) in an attempt to fit the properties of the PPNe HD 56126 (IRAS 07134$+$1005).  They computed models in a range of masses, 0.6$-$0.8 M$_\sun$ and luminosities, 4500$-$8000 L$_\sun$.  Several correlations were found: the period and amplitude of the pulsation decreased with increasing temperature and mass, and decreased with decreasing metallicity.  However, a good fit was not obtained for reasonable parameters.

Recently \citet{fad19} published models that also included convection.  The study examined temperatures in the range of 3000$-$15,000 K.  At temperatures greater than 4000 K, the pulsation is driven by $\kappa$-mechanism operating in the helium ionization zone.  The evolved masses were in the range of 0.54$-$0.62 M$_\sun$ and population I, solar abundances were used.
The general results were that the periods decreased with increasing temperature and decreasing mass, the latter in contradiction to the results of \citet{fok01}.  The amplitude was found to decrease with increasing temperature.  

The stars in our present study have temperatures in the range of 7000$-$9500 K and periods in the range of 24$-$73 days.  The periods calculated by \citet{fad19} are in the range 10$-$30 days at these temperatures, somewhat shorter than those observed.  Comparing the \citet{fad19} results to the larger sample of stars in Figure~\ref{PVT_plots}c, we find that the periods are a factor of 2$-$3 smaller than those observed.
Moreover, these calculations were for solar abundances; the stars in our study have an average [Fe/H] = -0.4.  \citet{fok01} found that the period decreased at lower metal abundance.
Thus we see that there are some pulsation models that attempt to fit the observed pulsation parameter of PPNe, but the agreement is not good and cannot be used to derive the mass of the PPNe.  Apparently, further work remains to be done on the modeling so that the potential contribution of these observational pulsation studies can be exploited to compare the mass and luminosity with stellar evolutionary models.

\subsection{Unusual Objects: IRAS 14429$-$4539 and 15482$-$5741}

IRAS 15482$-$5741 appears to have too low a luminosity be be a post-AGB object. 
One might consider it instead to be a star whose evolution was interrupted while on the red giant branch (RGB), which would explain its lower luminosity. 
A number of such post-RGB have been found in the Large and Small Magellanic Clouds \citep{kam14,kam15,kam16} at well-established distances.
However, evidence against this interpretation is the strong overabundance of {\it s}-process elements \citep{per12}, which are formed by neutron capture during the AGB phase of evolution.
Also, this object displays the 21 $\mu$m feature that has only been seen thus far in post-AGB objects and in particular in PPNe.
This object is unusual in that it displays evidence of a carbon chemistry in its circumstellar dust, with the presence of PAH features and the 21 $\mu$m feature \citep{cer11}, but an oxygen-rich chemistry according to its photospheric abundances \citep{per12}. 
This is the opposite of the expectation for normal stellar evolution of a star with a mixed chemistry.  Rather, one would  expect the star to begin its evolution and mass loss as an oxygen-rich star, and then later, following third dredge up on the AGB, to convert to a carbon-rich photosphere.\footnote{Mixed-chemistry planetary nebulae of low mass have been observed in the galactic bulge, particularly in objects with a dense torus.  This chemistry has been attributed to the formation of hydrocarbons in the gas-phase reactions in the dense disk, stimulated by ultraviolet irradiation \citep{guz11,guz14}.  However, this model is not relevant to RAS 15482$-$5741, which contains a central star of only 7400 K.}
This causes us to wonder about the possible close alignment of two different stars contributing to this unusual situation.
But even then, the visible star, showing the overabundance {\it s}-process elements, and the cool dust, showing the 21 $\mu$m feature, both possess the properties of highly evolved, and thus very luminous objects, while the composite SED is under-luminous for a post-AGB object.  
A perhaps more reasonable explanation for the low luminosity is an error in the Gaia distance.  The Gaia DR2 distance \citep{bai18} was significantly larger than that of the EDR3 value, with {\it D} = 8.7 kpc, resulting in {\it L} = 1600 {\it L}$_{\sun}$ using the average from the lower and upper extinction limits and as high as 3300 {\it L}$_{\sun}$ using the higher extinction value and the upper limit on the distance uncertainty.  
However, even if this is the case, it does not explain the reason for the unusual mixed chemistry. 
One might appeal to earlier mass transfer from a more evolved, carbon-rich companion to the oxygen-rich star.  These are speculations.

IRAS 14429$-$4539 is unusual in that it does not possess the clear bimodal SED seen in the other eight objects and typically seen in PPNe.  Its SED was fit better by three black body curves, although, as noted earlier (Sec.~\ref{seds}), the SED in the visible and near-IR could be explained by a large {\it A}$_{\rm v}$ due to circumstellar extinction.
Even then there was excess flux seen in the 2$-$5 $\mu$m region.  This suggests hot dust in the system, with a temperature of a few thousand K, perhaps resulting from a recent episode of mass loss in the system.  
Another alternative is a cool, low mass companion.  Since the near-infrared excess amounts to $\sim$100 {\it L}$_{\sun}$, this would imply that the companion is an evolved red giant.  This seems less likely, since it would imply that the original masses of the two stars were rather similar, since post-main sequence evolution occurs relatively rapidly.
This was the only object in the study that did not show evidence for periodic light variability.

\subsection{Multi-Year Periodicity $-$ Evidence of Binarity and/or Disks?}

Evidence of long-term, multi-year periodic variations are seen in three of these objects, IRAS 08143$-$4406, 14325$-$6428, and 15482$-$5741.  
These range from 5 to 19 years, although, as mentioned previously, the longest of these periods should be confirmed by additional observations since it is on the same order as the observing interval.
While these variations are seen most clearly in the ASAS-3 data, with its long-term coverage, they are also found in the combined SARA and ASAS-SN data for IRAS 08143$-$4406 and 15482$-$5741.  
The analyses of these light curves result in semi-amplitudes ranging from 0.02 to 0.10 mag, and they yield good fits to all three light curves.

One might initially consider the cause of these variations to be the oblate shape of the PPNe caused by the tidal interaction of a companion star.  However, as was discussed briefly in our previous paper in this series \citep{hri20b}, in which we found evidence of periods of 3.5 and 4.2 years in two of the objects, variations of even 0.02 mag in semi-amplitude are too large to be produced in this way.  Here, as in that paper, we suggest a more likely explanation as due to a partial obscuration by a circumbinary disk or dusty cloud as the PPNe orbits around the center of mass with an unseen companion.
This appears to indicate that long-term periodic variations and possible binary companions and circumbinary disks are more common in PPNe than previously realized.
\citet{sos21} recently presented strong evidence that the long secondary periods seen in red giants and AGB stars are due to binarity and obscuration by an associated dust cloud.  
What we are seeing here may be the same phenomenon.
In a forthcoming paper, we study in more detail several additional objects which show these long-term, periodic photometric variations (Hrivnak et al., in preparation).

\section{RESULTS AND SUMMARY}
\label{summary}

We carried out photometric {\it VR}$_C$ monitoring of nine Southern Hemisphere PPNe candidates over an eight-year interval to search for periodic light variations.  These were supplemented by more recent, high density observations with the ASAS-SN program, and in some cases by lower-precision observations made as part of the ASAS-3 program, both of which are publicly available.  These extended the observing interval by up to 10 additional years. These light curves were analyzed to search for periodicities in the data.
We list below the primary results of this study. 

1. Seven of the nine objects in this study are found to vary periodically, with periods ranging from 24 to 73 days, although the shorter of these are less well established.  The variations in brightness are not large, with maximum values ranging from 0.11 to 0.23 mag {\it V} in a season.

2. Luminosities of the objects range from 600 to 10,000  {\it L}$_\sun$ for the seven objects with Gaia distances and reasonably well-established extinction values, with uncertainties of $\sim$30 $\%$.  Five of them have values which agree well with those expected for PPNe; one other is consistent, within errors, with the values of PPNe but one is much too low.  

3. The pulsational properties of these objects agree well with the general trends seen in previous studies of the variability of PPNe, showing correlations between period, maximum seasonal brightness variation, and {\it T}$_{\rm eff}$.
The periods and maximum brightness variations are smaller at higher temperatures.

4. Secondary periods are seen in the four objects with the longest primary periods, and the ratios ({\it P}$_2$/{\it P}$_1$) range from 0.8 to 1.1.
These are in the range of values found in previous studies of PPNe.

5. The light curves of three of the objects show long-term, multi-year periodicities with values ranging from 5 to 19 years.  These are suggested to be cause by the periodic occultation of the star by a circumbinary disk as it orbits with its binary companion. 

6. Evidence in the literature indicates that IRAS 15482$-$5741 has an O-rich photosphere, an over-abundance of {\it s}-process elements, and a C-rich circumstellar dust envelope. This object also has the low luminosity of 600 {\it L}$_\sun$.  These unusual combination of properties commends further investigation of this object.

We now possess a reasonable sample of approximately 30 PPNe for which we have periods, measured seasonal brightness variations, and {\it T}$_{\rm eff}$, or for which we know at least two of these properties. 
However, there are not stellar pulsation models available that provide a good to match these observed properties.  Thus we put forth an invitation and appeal for modelers to address this situation.
Successful pulsation models can yield the masses of the objects, and combined with even more precise distance measurement from Gaia will yield luminosity values which will allow one to test and constrain theoretical mass-luminosity models of post-AGB stellar evolution.
Radial velocity studies of the three objects with long-term periodicity would be a good way to investigate the suggestion of their binarity.  At present, only a few PPNe, such as IRAS 08005$-$2356 \citep{man21}, have been found to be in binary systems.  

\acknowledgments

We thank the anonymous referee for helpful comments and for pointing to some particularly relevant references.
BJH acknowledges ongoing support from the National Science Foundation (1413660) and the Indiana Space Grant Consortium.
The ongoing technical support of Paul Nord and the use of Kevin Volk's SED program are gratefully acknowledged.
This research has made use of the SIMBAD and VizieR databases, operated at CDS, Strasbourg,
France, NASA's Astrophysical Data System, and Peranso (www.peranso.com), a light curve and period analysis software.

\facility{SARA}
\facility{ASAS-3}
\facility{ASAS-SN}

\clearpage

\tablenum{1}
\begin{deluxetable}{ccrrrccccl}
\rotate
\tablecaption{List of PPN Targets Observed \label{object_list}}
\tabletypesize{\footnotesize} \tablewidth{0pt} \tablehead{
\colhead{IRAS ID}&\colhead{GLMP ID\tablenotemark{a}}&\colhead{2MASS ID}&\colhead{R.A.\tablenotemark{b}}&\colhead{Decl.\tablenotemark{b}}
&\colhead{{\it l}}&\colhead{{\it b}}
&\colhead{{\it V}\tablenotemark{c}}&\colhead{{\it B$-$V}\tablenotemark{c}}&\colhead{Sp.T.} \\
&&&(2000.0)&(2000.0)&($\arcdeg$)&($\arcdeg$)&\colhead{(mag)} &\colhead{(mag)} & } 
\startdata
08143$-$4406 & 206 & 08160303-4416045 & 08:16:03.03 & $-$44:16:04.5 & 260.8 & $-$05.1 & 12.3  & 1.5 & F8~I\tablenotemark{d}, K1-2~I\tablenotemark{e} \\
08281$-$4850 & 218 & 08294055-4900043 & 08:29:40.55 & $-$49:00:04.3 & 266.1 & $-$05.8 & 13.9  & 1.6 & F0~I\tablenotemark{e} \\
11387$-$6113 & 317  & 11410870-6130173 & 11:41:08.70 & $-$61:30:17.3 & 294.6 & $+$00.2 & 11.8  & 1.1 & A3~Ie\tablenotemark{e} \\
12360$-$5740 & 334 & 12385310-5756318 & 12:38:53.10 & $-$57:56:31.8 & 301.3 & $+$04.9 & 12.4  & 1.1 & \nodata \\ 
13245$-$5036 & 347 & 13273694-5052061 & 13:27:36.94 & $-$50:52:06.1 & 308.7 & $+$11.6 & 12.3 & 0.8 & A7-9~Ie\tablenotemark{e}  \\
14325$-$6428 & \nodata & 14363437-6441310 & 14:36:34.37 & $-$64:41:31.1 & 313.9 & $-$04.1  & 11.8  & 1.0 & F5~I\tablenotemark{e} \\
14429$-$4539 & \nodata & 14461377-4552051 & 14:46:13.77 & $-$45:52:05.1 & 323.0 & $+$12.5 & 13.5 & 0.9 & F4~Ie\tablenotemark{e}, G0~Ie\tablenotemark{f} \\
15310$-$6149 & \nodata & 15351712-6159041 & 15:35:17.12 & $-$61:59:04.1 & 320.9 & $-$05.0 & 13.6 & 0.8 & A7~I\tablenotemark{e}, F0~Ve\tablenotemark{g} \\
15482$-$5741\tablenotemark{h} & \nodata & 15521945-5750531 & 15:52:19.45 & $-$57:50:53.2 & 325.2 & $-$03.0 & 13.5 & 1.0 & F7~I\tablenotemark{e} \\
\enddata
\tablenotetext{a}{\citet{garlar97}.}
\tablenotetext{b}{Coordinates from the 2MASS Catalog.}
\tablenotetext{c}{These values are all variable as discussed in this paper.  They are based on our measurements, except for the {\it B}$-${\it V} measurements of IRAS 11387$-$6113, 14429$-$4539 \citep{hen12}, and 15482$-$5741 \citep{nis16}.} 
\tablenotetext{d}{Spectral type by \citet{red96}. } 
\tablenotetext{e}{Spectral types by \citet{suarez06}. } 
\tablenotetext{f}{Spectral type by \citet{hu93}. }
\tablenotetext{g}{Spectral type by \citet{vie03}. }
\tablenotetext{h}{The {\it IRAS} position is off by 5$\arcsec$ to the SE; however, there is good agreement between the 2MASS position and those measured for the infrared source with {\it WISE} and {\it AKARI}.} 
\end{deluxetable}

\tablenum{2}
\begin{deluxetable}{lrrrrl}
\tablecaption{Observed Standard Magnitudes and Colors of the Program Stars
\label{std_ppn}}
\tabletypesize{\footnotesize} 
\tablewidth{0pt} \tablehead{ \colhead{IRAS ID} &\colhead{{\it V}}
 &\colhead{{\it B$-$V}} &\colhead{{\it V$-$R$_C$}} &\colhead{{\it V$-$I$_C$}} 
&\colhead{Date}  \\
&\colhead{(mag)} & \colhead{(mag)} & \colhead{(mag)} & \colhead{(mag)} 
&\colhead{} } 
\startdata
08143$-$4406  & 12.19   & 1.49 & 0.86 & 1.75 & 1993 Apr 18\tablenotemark{a}   \\ 
			& 12.41   & \nodata & 0.90 & \nodata &  2014 Jun 03 \\ 
08281$-$4850   & 13.96  & 1.55 & 0.98  & 1.99 & 1993 Apr 18\tablenotemark{a}   \\ 
			 & 13.86   & \nodata & 0.94 & \nodata &  2014 Jun 03 \\ 
11387$-$6113 & 11.75   & \nodata & 0.69\tablenotemark{b} & \nodata &  2013 Jul 17 \\ 
			& 11.76   & \nodata & 0.68\tablenotemark{b} & \nodata &  2014 Jun 03 \\ 
12360$-$5740  & 12.48 & 1.14 & 0.66  & 1.30 & 1992 May 19\tablenotemark{a} \\ 
			& 12.39   & \nodata & 0.64 & \nodata &  2013 Jul 17 \\ 
			& 12.38   & \nodata & 0.62 & \nodata &  2013 Aug 21 \\ 
			& 12.37   & \nodata & 0.61 & \nodata &  2014 Jun 03 \\ 
13245$-$5036	& 12.28 & 0.77 & 0.44\tablenotemark{b}  & 0.86\tablenotemark & 1992 May 19\tablenotemark{a} \\ 
			& 12.38   & \nodata & 0.42\tablenotemark{b} & \nodata &  2013 Jul 17 \\  
			& 12.31   & \nodata & 0.43\tablenotemark{b}& \nodata &  2013 Aug 21 \\ 
			& 12.36   & \nodata & 0.43\tablenotemark{b} & \nodata &  2014 Jun 03 \\ 
14325$-$6428	& 11.76 & 1.05 & 0.66  & 1.39 & 1992 May 19\tablenotemark{a} \\  
			& 11.84   & \nodata & 0.67 & \nodata &  2013 Jul 17 \\ 
			& 11.78   & \nodata & 0.66 & \nodata &  2014 Jun 03 \\ 
14429$-$4539  & 13.53   & \nodata & 0.65\tablenotemark{b} & \nodata &  2014 Jun 03 \\ 
15310$-$6149   & 13.60  & 0.84 & 0.54  & 1.11 & 1992 May 19\tablenotemark{a} \\  
			& 13.55   & \nodata & 0.52 & \nodata &  2013 Jul 17 \\ 
			& 13.55   & \nodata & 0.54 & \nodata &  2013 Aug 21 \\ 
			& 13.55   & \nodata & 0.53 & \nodata &  2014 Jun 03 \\ 
15482$-$5741  & 13.51   & \nodata & 0.86 & \nodata &  2014 Jun 03 \\ 
\enddata
\tablecomments{Uncertainties in the brightness and color are $\pm$0.01$-$0.02 mag.}
\tablenotetext{a}{Observed earlier at Cerro Tololo Interamerican Observatory.}
\tablenotetext{b}{Note that the {\it R}$_C$ measurement contains a contribution from H$\alpha$ emission.}
\end{deluxetable}

\tablenum{3}
\begin{deluxetable}{llrrr}
\tablecaption{Comparison Star Identifications, Standard Magnitudes and Colors\label{std_comp}}
\tabletypesize{\footnotesize}
\tablewidth{0pt} \tablehead{\colhead{IRAS Field}
&\colhead{Star} &\colhead{2MASS ID} &\colhead{{\it V}} 
&\colhead{{\it V$-$R}$_C$}  \\
\colhead{}
&\colhead{} &\colhead{}  &\colhead{(mag)} 
 &\colhead{(mag)}} \startdata
08143$-$4406 & C1 & 08155545$-$4415131 & 12.53 & 0.71  \\
              & C2 & 08160306$-$4418063 & 12.66 & 0.31  \\
              & C3 & 08160356$-$4414480 & 13.26 & 0.63  \\
08281$-$4850 & C1 & 08292761$-$4900067 & 14.04 & 0.76  \\
              & C2 & 08293981$-$4859526 & 13.00 & 0.24  \\
              & C3 & 08293499$-$4902039 & 13.93 & 1.06  \\
11387$-$6113 & C1 & 11410730$-$6132349 & 12.58 & 0.68  \\
              & C2 & 11404932$-$6129288 & 13.40 & 0.68  \\
              & C3 & 11404633$-$6130121 & 13.02 & 0.40  \\
12360$-$5740 & C1 & 12385047$-$5759449 & 12.80 & 0.45  \\
              & C2 & 12383218$-$5753165 & 12.74 & 0.69  \\
              & C3 & 2391486$-$5800160 & 12.77 & 0.32  \\
13245$-$5036 & C1 & 13272167$-$5055377 & 12.41 & 0.34  \\
              & C2 & 13272006$-$5055090 & 12.24 & 0.61  \\
              & C3 & 13274885$-$5050174 & 12.07 & 0.62  \\
14325$-$6428 & C1 & 14370825$-$6439179 & 11.95 & 1.11  \\
              & C2 & 14371064$-$6439377 & 12.55 & 1.20  \\
              & C3 & 14364940$-$6438209 & 12.71 & 0.98  \\
14429$-$4539 & C1 & 14462277$-$4552243 & 12.94 & 0.75  \\
              & C2 & 14461036$-$4551148 & 13.06 & 0.56  \\
              & C3 & 14460267$-$4552437 & 12.69 & 0.67  \\
15310$-$6149 & C1 & 15352765$-$6159054 & 13.06 & 0.44  \\
              & C2 & 15351855$-$6159060 & 12.68 & 0.26  \\
              & C3 & \nodata\tablenotemark{a} & 13.04 & 0.32  \\
15482$-$5741 & C1 & 15522431$-$5752318 & 12.86 & 0.84  \\
              & C2 & 15524149$-$5751428 & 13.11 & 1.10  \\
              & C3 & 15522020$-$5750277 & 13.66 & 1.19  \\
\enddata
\tablecomments{Uncertainties in the brightness and color are $\pm$0.01$-$0.02 mag. }
\tablenotetext{a}{Not in 2MASS Catalog; Gaia DR2 5827420393293216384.}
\end{deluxetable}

\tablenum{4}
\begin{deluxetable}{lcrrrrrrrrrrrrr}
\tablecolumns{17} \tabletypesize{\scriptsize}
\tablecaption{Main Results of the Periodogram Study of the Light Curves\tablenotemark{a,b}\label{periods}}
\rotate
\tabletypesize{\footnotesize} 
\tablewidth{0pt} \tablehead{ 
\colhead{IRAS ID} &\colhead{Filter} & \colhead{Years} & \colhead{Data\tablenotemark{c}} & \colhead{No.} & \colhead{{\it P}$_1$} & \colhead{{\it A}$_1$} & \colhead{{\it $\phi$}$_1$\tablenotemark{d}} & \colhead{{\it P}$_2$}&\colhead{{\it A}$_2$} &\colhead{$\phi$$_2$\tablenotemark{d}} & \colhead{{\it P}$_3$} & \colhead{{\it A}$_3$} & \colhead{$\phi$$_3$\tablenotemark{d}} & \colhead{$\sigma$\tablenotemark{e}} \\
 & & & \colhead{Sets} &\colhead{Obs.} & \colhead{(days)}&\colhead{(mag)} & &\colhead{(days)}&\colhead{(mag)}
 & &\colhead{(days)} &\colhead{(mag)}&  &\colhead{(mag)} }
\startdata
\multicolumn{14}{c}{Pulsation Results} \\
\tableline 
08143$-$4406 & {\it V}\tablenotemark{f} & 2010-2018 & 2,3 & 273 & 57.4 & 0.040 & 0.56 & 76.3 & 0.034 & 0.68 & 78.4 & 0.027 & 0.76  & 0.026 \\
08281$-$4850 & {\it V}\tablenotemark{f} & 2010-2018 & 2 & 73 & 48.4 & 0.032 & 0.64 & 55.6 & 0.029 & 0.10  & \nodata & \nodata & \nodata & 0.021 \\
11387$-$6113 & {\it V}\tablenotemark{f} & 2010-2018 & 2,3 & 293 & 37.4 & 0.029 & 0.07 & \nodata & \nodata & \nodata  & \nodata & \nodata & \nodata & 0.035 \\
12360$-$5740 & {\it V}\tablenotemark{f} & 2010-2018 & 2,3 & 302 & 44.8 & 0.023 & 0.93 & 48.2 & 0.018 & 0.80 & 53.6 & 0.015 & 0.51 & 0.025 \\
13245$-$5036  & {\it V}\tablenotemark{f} & 2010-2018 & 2,3 & 316 & 23.9 & 0.030 & 0.61 & 20.4: & 0.018 & 0.58  & \nodata & \nodata & \nodata & 0.035 \\
14325$-$6428 & {\it V}\tablenotemark{f} & 2010-2018 & 2,3 & 266 & 50.9 & 0.017 & 0.97 & 46.9 & 0.017 & 0.49  & \nodata & \nodata & \nodata & 0.024 \\
15310$-$6149 & {\it R}$_C$\tablenotemark{f} & 2010-2018 & 2 & 52 & 37.3 & 0.008 & 0.15 & \nodata & \nodata & \nodata & \nodata & \nodata & \nodata & 0.008 \\
\tableline 
\multicolumn{14}{c}{Long-Period Results} \\
\tableline 
08143$-$4406 & {\it V} & 2000-2018 & 1,2,3 & 858 & 5.0 yr & 0.021 & 0.79 & \nodata & \nodata & \nodata & \nodata & \nodata & \nodata & 0.051 \\
14325$-$6428& {\it V} & 2000-2018 & 1,2,3 & 790 & 18.8 yr & 0.096 & 0.41 & \nodata & \nodata & \nodata & \nodata & \nodata & \nodata & 0.041 \\
15482$-$5741 & {\it V} & 2000-2018 & 1,2,3 & 583 & 9.6 yr & 0.037 & 0.93 & \nodata & \nodata & \nodata & \nodata & \nodata & \nodata & 0.053 \\
\enddata
\tablenotetext{a}{The uncertainties in the parameters are as follows: period ({\it P}) $-$ $\pm$0.01$-$0.1 days for pulsation and $\pm$0.1$-$0.2 yr for the long-periods; amplitude ({\it A}) $-$  $\pm$0.002$-$0.004 mag; phase ($\phi$)  $-$ $\pm$0.01$-$0.03. }
\tablenotetext{b}{Colons (:) indicate less certain period values or ones that fell slightly below our adopted level of significance; see text for details.}
\tablenotetext{c}{1 = ASAS-3, 2 = SARA, 3 = ASAS-SN.}
\tablenotetext{d}{The phases are determined based on the epoch of 2,455,600.0000, and they each represent the phase derived from a sine-curve fit to the data, not the phase of minimum light.}
\tablenotetext{e}{Standard deviation of the observations from the sine-curve fit.}
\tablenotetext{f}{Analysis based on the seasonally normalized light curve.}
\end{deluxetable}

\tablenum{5}
\begin{deluxetable}{rrrrcrrcl}
\tablecaption{Calculated Luminosities of the Program Objects\tablenotemark{a}
 \label{sed_table}}
\tabletypesize{\footnotesize}
\tablewidth{0pt} \tablehead{ \colhead{IRAS ID} & \colhead{{\it Gaia} DR2 ID\tablenotemark{b}} &\colhead{{\it D}\tablenotemark{c}} &\colhead{{\it A}$_V$(C)} 
&\colhead{{\it A}$_V$(SF)}  &\colhead{{\it L}(C)\tablenotemark{d}} &\colhead{{\it L}(SF)\tablenotemark{d}} & Chem & \colhead{Type} \\
\colhead{} &\colhead{} &\colhead{(kpc)} &\colhead{(mag)}&\colhead{(mag)} &\colhead{(L$_{\sun}$)} &\colhead{(L$_{\sun}$)} &\colhead{} &\colhead{} }
\startdata
08143$-$4406 & 5520238967817034880 & 4.16 & 0.58 & 2.84 & 1400 & 2800 & C & post-AGB \\
08281$-$4850 & 5515266327706463616 & 11.45 & 0.74: & 2.37 &9200: & 11,200 & C & post-AGB \\
11387$-$6113 & 5335675087769798272 & 5.02 & 0.92 & 3.00:\tablenotemark{e} & 3500 & 6600:\tablenotemark{e} & O\tablenotemark{f} & post-AGB \\
12360$-$5740 & 6060828565581083264 & 9.08 & 0.95 & 1.58 & 3500 & 4300 & O & post-AGB\\
13245$-$5036 & 6070128028770373888 & 14.21 & 0.30: & 0.61 & 7000: & 7600 & C & post-AGB\\ 
14325$-$6428 & 5849962851220246016 & 4.88 & 1.33: & 1.70 & 4100: & 4400 & C & post-AGB \\
14429$-$4539 & 5906408788891928704 & 3.85 & 0.36 & 0.43 & 4100 &  4100  & C & post-AGB \\
15310$-$6149 & 5827431968181256576 & \nodata & 1.00: & 1.43  &  \nodata  & \ \nodata  & C: & post-AGB \\
15482$-$5741 & 5835842751094231808 & 5.55 & 1.31: & 1.96 & 500: & 700 & O+C & post-RGB\\
\enddata
\tablecomments{{\it A}$_V$(C) and {\it A}$_V$(SF) refer to the {\it V} extinction as determined from \citet{cap17} and \citet{sf11}, respectively.  
{\it L}(C) and {\it L}(SF) refer to the luminosities determined with the \citet{cap17} and with the \citet{sf11} extinction values, respectively.
Note that these values of {\it A}$_V$ are in good agreement with those from \citet{vic15} for the three objects in common. }
\tablenotetext{a}{The colons indicate more uncertain values.}
\tablenotetext{b}{Gaia EDR3 = DR2 ID.}
\tablenotetext{c}{Uncertainties in the distances typically range from a lower limit of $\sim$0.9 and an upper limit of $\sim$1.2 times the distance listed.  }
\tablenotetext{d}{The uncertainties in the luminosities, based on the uncertainties in the distances, typically range from a lower limit of a factor of $\sim$0.8 to an upper limit of a factor of $\sim$1.4 for each of the objects. } 
\tablenotetext{e}{For IRAS 11387$-$6113, {\it A}$_V$(SF) = 6.95 mag, a value much larger than for any of the other objects.  This leads to {\it L}(SF) = 130,000 L$_{\sun}$, an extremely large luminosity.  These appear to be much too large for a post-AGB object.  For the purpose of getting an approximate value for {\it L}, we adopted  {\it A}$_V$(SF) = 3.00.}   
\tablenotetext{f}{Perhaps mixed chemistry, as there is evidence of PAH features in the mid-IR spectrum \citep{ram17}.}
\end{deluxetable}

\tablenum{6}
\begin{deluxetable}{rrrcrcccrr}
\tablecaption{Results of Our Period and Light Curve Study\label{results}} 
\tabletypesize{\footnotesize}
\tablewidth{0pt} \tablehead{ \colhead{IRAS ID} &\colhead{{\it V}} &\colhead{$\Delta${\it V}\tablenotemark{a}} 
&\colhead{SpT} &\colhead{{\it T}$_{\rm eff}$\tablenotemark{b}}  &\colhead{Chem} &\colhead{{\it P}$_1$} &\colhead{{\it P}$_2$} &\colhead{{\it P}$_2$/{\it P}$_1$} & \colhead{P(long)}\\
\colhead{} &\colhead{(mag)} &\colhead{(mag)} &\colhead{} &\colhead{(K)} &\colhead{} &\colhead{(days)} &\colhead{(days)} &\colhead{} & \colhead{(yrs)}}
\startdata
08143$-$4406 &  12.3 & 0.23 & F8~I & 7125 & C & 73 &58 & 0.79 & 5.0 \\
08281$-$4850 &  13.9 & 0.13 & F0~I & 7875 & C & 48 & 56 & 1.15 & \nodata  \\
11387$-$6113\tablenotemark{c}  &   11.8 & 0.19 & A3~Ie & \nodata & O\tablenotemark{d} & 37: & \nodata & \nodata & \\
12360$-$5740 &  12.4 & 0.16 & \nodata & 7400 & O & 45 & 48 & 1.08 & \nodata\\
13245$-$5036 &  12.3 & 0.20 & A7-9~Ie & 9500 & C & 24: & \nodata & \nodata & \nodata  \\
14325$-$6428 &  11.8 & 0.14 & F5~I & 8000 & C & 51 & 47 & 0.91 & 18.8 \\
14429$-$4539 &  13.5 & 0.11 & F4~Ie, G0~Ie & 9375 & C & \nodata & \nodata & \nodata & \nodata \\
15310$-$6149 &  13.6 & 0.16 & A7~I & \nodata & C: & 37: & \nodata & \nodata & \nodata \\
15482$-$5741 &  13.5 & 0.15 & F7~I & 7400 & O+C & \nodata & \nodata & \nodata & 9.6 \\
\enddata
\tablenotetext{a}{The maximum brightness range observed in a season.}
\tablenotetext{b}{Temperatures from high-resolution spectral observations and abundance analyses: IRAS 08143$-$4406, 08281$-$4850, 13245$-$5036, 14325$-$6428, and 14429$-$4539 $-$ \citet{des16}; IRAS 12360$-$5740 and 15482$-$5741 $-$ \citet{per12}. }
\tablenotetext{c}{Included in the ASAS-SN Variable Star Catalog as J114108.71-613017.4, {\it P}=37.63 days. }
\tablenotetext{d}{Perhaps mixed chemistry, as there is evidence of PAH features in the mid-IR spectrum \citep{ram17}.}
\end{deluxetable}

\clearpage

\clearpage

\clearpage

\clearpage

\clearpage

\clearpage

\end{document}